\documentclass[
preprint,
  nofootinbib,amsmath,amssymb,aps,preprintnumbers,superscriptaddress,notitlepage
]{revtex4-1}

\usepackage{graphicx}
\usepackage{bm}
\usepackage{braket}
\usepackage{cases} 
\usepackage{here}
\usepackage{color}
\usepackage{upgreek}
\usepackage{ulem}
\usepackage{comment}
\usepackage{empheq}

\def\la{\langle}
\def\ra{\rangle}
\newcommand{\be}{\begin{equation}}
\newcommand{\ee}{\end{equation}}
\newcommand{\eq}[1]{(\ref{#1})}
\def\bea{\begin{eqnarray}}
\def\eea{\end{eqnarray}}

\begin{document}

\title{Gravitational Waves in Metastable Supersymmetry Breaking}

\author{Chong-Sun Chu}
\email[]{cschu@phys.nthu.edu.tw}
\affiliation{Department of Physics, National Tsing-Hua University,
  Hsinchu 30013, Taiwan}
\affiliation{Center for Theory and Computation, National Tsing-Hua
  University,  Hsinchu 30013, Taiwan}
\affiliation{Physics Division, National Center for Theoretical Sciences,
  Taipei 10617, Taiwan}

\author{Asuka Ito}
\email[]{asuka.i.aa@m.titech.ac.jp}
\affiliation{Department of Physics, Tokyo Institute of Technology,
  Tokyo 152-8551, Japan}
\affiliation{Department of Physics, National Tsing-Hua University,
  Hsinchu 30013, Taiwan}
\affiliation{Center for Theory and Computation, National Tsing-Hua
  University, Hsinchu 30013, Taiwan}

\date{\today}

\begin{abstract}
If supersymmetry is broken in metastable vacua, it is not clear why we
are now in there rather than supersymmetric vacua.  Moreover, it is
natural to expect that we were in supersymmetric vacua, which have
higher symmetry than metastable vacua, in the early universe.  In this
paper, we reexamine and improve the previous analysis on the
cosmological evolution of the vacuum structure in the ISS model of
metastable supersymmetry breaking by taking into account
constraints on the reheating temperature, which is needed to avoid the
overproduction of gravitinos.  It turns out that the desired phase
transition from a supersymmetric vacuum to a metastable vacuum is
allowed only in the light gravitino mass region $m_{3/2} < 4.7$\,eV.
This is achieved by either rolling down potential or tunneling
processes depending on the reheating temperature.  
We show that when the tunneling processes
are realized, abundant gravitational waves could be produced from collisions of runaway bubbles.
The resulting gravitational waves are detectable with the future
gravitational wave interferometers like LISA and DECIGO.
\end{abstract}

\maketitle

%
%
%
%
%
%
%
\newpage

\section{Introduction}
Gravitational waves are a powerful tool to explore the early universe.
Indeed, gravitational waves from 
inflation~\cite{Grishchuk:1974ny,Starobinsky:1979ty,Ito:2020neq}
have been probed intensively by
the temperature anisotropy and the B-mode polarization of the cosmic microwave 
background and
current null observations have put constraints on the energy scale of 
inflation~\cite{Planck:2018jri,BICEP:2021xfz}.
Recently, pulsar timing arrays, NANOGrav, PPTA, EPTA and IPTA reported
detection of correlated
signals among pulsars~\cite{NANOGrav:2020bcs,Goncharov:2021oub,Chen:2021rqp,Chen:2021ncc}, which might be
stochastic gravitational waves of 
primordial origin.  In particular, it has been discussed that primordial
gravitational waves from the (first order) QCD phase transition can
account for the pulsar timing
signals~\cite{Neronov:2020qrl,Li:2021qer}.  The electroweak phase
transition, if it is first order, can also result in abundant
gravitational
waves~\cite{Apreda:2001us,Grojean:2006bp,Kakizaki:2015wua,Ellis:2018mja},
which could be detected by gravitational wave interferometers such as
LIGO~\cite{LIGO}, Virgo~\cite{Virgo}, LISA~\cite{LISA}, and
DECIGO~\cite{DECIGO}.  Therefore, the observation of
gravitational waves 
enables us to explore high energy physics in the early universe and is a
promising way to
access physics that goes
beyond current collider experiments $\sim \mathcal{O}(1)\,$TeV.

Supersymmetry is often postulated to be a symmetry of the
fundamental theory at high energy.  However if it is true, then
supersymmetry needs to be broken at some point in the cosmological
history.
In~\cite{Craig:2020jfv}, it was pointed out that
a first order phase transition along a pseudo-flat direction associated
with the spontaneous breaking of a $U(1)$ R-symmetry could give rise to
detectable gravitational wave signals.
The discussion
applies to generic supersymmetry breaking models where the supersymmetry breaking
vacua are global minima,
since the existence and breaking of R-symmetry is usually
required in this case~\cite{Nelson:1993nf}.
An exception to the discussion is the metastable
supersymmetry breaking model proposed by Intriligator,
Seiberg, and Shih (ISS)~\cite{Intriligator:2006dd}, where
the metastable
supersymmetry breaking vacuum is local but not global minimum.
The model has attracted
substantial attention
because it can avoid the difficulties associated with 
R-symmetry~\cite{Nelson:1993nf} and
chirality~\cite{Witten:1982df}. Moreover the metastable vacuum
can be made to be parametrically long-lived, making the metastable supersymmetry breaking a
viable scenario.

Nevertheless, it is natural to expect that we were in
a supersymmetric vacuum in the early universe, which has higher symmetry, 
than a metastable vacuum.
It is then necessary to understand 
how we ended up in the metastable vacuum now. To answer the question,
\cite{Abel:2006cr,Abel:2006my,Fischler:2006xh,Craig:2006kx} 
studied the phase structure of the theory by taking into account
thermal corrections from particles in the ISS hidden sector
to the effective potential
and showed that the metastable vacuum could be
chosen eventually  in the cosmological evolution
of the thermal universe even if the universe was
in the supersymmetric vacuum initially.
However, it should be stressed that an arbitrary high reheating temperature
is allowed in the analysis
of these early works. 
On the other hand, in gauge mediated supersymmetry
breaking models, the reheating temperature actually
cannot be too high in order to
avoid the overproduction of gravitinos~\cite{Weinberg:1982zq}.

In this paper, we will first reexamine the previous analysis~\cite{Abel:2006cr,Abel:2006my}
in light of the gravitino problem.
We will find that a phase transition from the
supersymmetric vacuum to the metastable vacuum is not possible in the
middle $0.4 \,{\rm keV} \lesssim m_{3/2} \lesssim 1 \,{\rm GeV}$ and 
heavy $ 600 \,{\rm GeV} \lesssim m_{3/2}$ allowed regions~\cite{Hook:2018sai}
of gravitino mass. 
As a result, in order for the
universe to be able to arrive at the metastable vacuum, a
messenger model that is compatible with the light gravitino mass is needed.
We will consider the light gravitino mass region
$m_{3/2} < 4.7$\,eV~\cite{Hook:2018sai} 
and show that, as long as a compatible messenger sector is chosen,
the desired phase transition 
is always possible
by taking into account thermal corrections from the ISS hidden sector and 
the minimal supersymmetric standard model (MSSM) as a visible sector.
The phase transition is achieved by either rolling down potential or
tunneling processes 
depending on the reheating temperature.
It will be argued that inclusion of effects from a messenger sector does not
spoil the conclusion.
Next, we will study 
gravitational wave production from the tunneling processes.
It will turn out that from the tunneling, 
gravitational waves are produced mainly by collision
of nucleated runaway bubbles. 
We will calculate
the
gravitational wave spectra and 
show that they are detectable with
the
future space borne
gravitational wave interferometers
such as
LISA~\cite{LISA} and DECIGO~\cite{DECIGO}.
This gives us a unique way to probe the metastable supersymmetry breaking.

The organization of the paper is as follows.
In  section \ref{ISS}, we review the ISS model of the dynamical
supersymmetry breaking in metastable vacua.  
In  section \ref{secgra}, the cosmological gravitino problem is reviewed.  
To avoid their overproduction, we see that high reheating temperature
is forbidden for certain range
of gravitino mass.
In  section \ref{secT}, we investigate the thermal contributions from 
the ISS sector and the MSSM sector
to the finite temperature effective potential.
We show that phase transitions from the supersymmetric
vacuum to the metastable vacuum are always possible 
in the light gravitino mass region.
We argue that inclusion of contributions from a messenger sector does not
spoil our conclusion.
In section \ref{GW}, we study gravitational wave production from
tunneling associated with the phase transition.
It is shown that the produced
gravitational waves are detectable with the future gravitational wave
interferometers like LISA~\cite{LISA} and
DECIGO~\cite{DECIGO}.  
The final section is devoted to conclusion and further discussions.
\section{Metastable vacua in supersymmetric gauge theory} \label{ISS}

For the sake of establishing
the notations, let us briefly
review the metastable supersymmetry breaking model
proposed by Intriligator, Seiberg, and Shih
(ISS)~\cite{Intriligator:2006dd}.  We consider a supersymmetric gauge
theory as a hidden sector for  supersymmetry breaking with chiral
superfields $\Phi_{ij}$, $\varphi^{i}_{c}$, and $\tilde{\varphi}^{i
  c}$ and a tree level superpotential
\begin{equation}
  W_{\rm cl} = h {\rm Tr} \varphi \Phi \tilde{\varphi} -
  h \mu^{2} {\rm Tr}\Phi \ , \label{supo}
\end{equation}
where $i=1\dots N_{f}$ denotes flavors, $c=1\dots N$ denotes
indexes of the $SU(N)$ gauge group, and $h$ and $\mu$ are constants. 
The number of flavors is taken to satisfy the condition $N_f  > 3N$
so that the theory is weakly coupled at energy scale much less than
the Landau scale $\Lambda_m$. 
We mention that above the energy
scale of the Landau pole, we have a UV free electric dual description
in terms of a $SU(N_f-N)$ gauge theory thanks to the Seiberg's
duality~\cite{Seiberg:1994pq,Intriligator:1995au,Peskin:1997qi}.  
Supersymmetry is broken in the ISS model
since the
$F$-flatness condition
cannot
be satisfied due to the rank condition.
The supersymmetry breaking vacuum has 
a non-vanishing energy density $V_{\rm meta}$ and is given by
\begin{equation}
  \la\Phi\ra = 0 \ ,
  \quad \la \varphi_c^i\ra = \la\tilde{\varphi}^{ic}\ra =  \mu
                      \delta_{ci}\ , \quad 
           V_{\rm meta} = \left( N_{f} -N \right) |h^{2} \mu^{4}| \ ,
           \label{brevacu}
\end{equation}
where $c,i =1, \cdots, N$
in a suitable choice of bases.
The vacuum is locally stable
\cite{Intriligator:2006dd} due to the 
Coleman-Weinberg  one-loop
contributions to the 
potential \cite{Coleman:1973jx}.
However it is not globally stable because there exists a supersymmetric vacuum, which is generated dynamically.

To see this, let us consider the situation that $\Phi$ has a non-zero vacuum expectation value.
Then from the superpotential (\ref{supo}), $\varphi$ and $\tilde{\varphi}$ obtain a mass of $h\Phi$.
Below the mass scale, one can integrate out $\varphi$ and $\tilde{\varphi}$ so that we have a pure supersymmetric $SU(N)$ gauge theory effectively.
The low energy effective theory is asymptotic free and we expect gaugino condensation.
Actually, a
superpotential is generated dynamically
\be \label{supo2}
W_{\rm dyn} = N \left(\frac{h^{N_{f}}
    {\rm det}\Phi}{\Lambda_{m}^{N_{f}-3N}} \right)^{1/N} ,
\ee
and  gives rise to the gaugino condensation~\cite{Intriligator:1995au,Intriligator:2007cp} 
\begin{equation}
  \la {\rm Tr}\rho \rho \ra
   = 32 \pi^{2} \left(
   \frac{h^{N_{f}} {\rm det}\Phi}{\Lambda_{m}^{N_{f}-3N}} \right)^{\frac{1}{N}} ,
                    \label{conden}
\end{equation}
where $\rho$ denotes
the gauginos. 
The gaugino condensate 
sets the mass scale for the massive $SU(N)$ degrees of
freedom such as the gauginos and the gauge fields of the ISS model
and will play a role in our discussion later for 
the finite temperature effective potential.
Moreover, the superpotential $W_{\rm dyn}$ leads to the emergence of a supersymmetry preserving
stable vacuum at
\begin{equation}
  \la \Phi \ra = \Phi_{0} \delta_{ij} \equiv
  \frac{\mu}{h \epsilon^{(N_{f}-3N)/(N_{f}-N)}} \delta_{ij} \ ,  
  \quad
  \la \varphi \ra =
  \la \tilde{\varphi}^{T} \ra =  \bm{0} \ ,  \label{suvacu}
\end{equation}
where $\epsilon := \frac{\mu}{\Lambda_{m}} \ll 1$.
We require $\epsilon \ll 1$ to guarantee the perturbative treatment of
the theory and
the longevity of the metastable vacuum (\ref{brevacu})~\cite{Intriligator:2006dd}.
Then the theory realizes supersymmetry breaking, assuming that 
we have already been in the metastable vacuum.

However it is actually more natural to  expect that we were in the
supersymmetric vacuum which has higher symmetry 
than the metastable vacuum in the early universe.
In \cite{Abel:2006cr,Abel:2006my,Fischler:2006xh,Craig:2006kx}, 
thermal
correction
from particles in the ISS sector to the structure
of vacua
was studied and
it was shown that the metastable vacuum could be preferred in the
thermal universe with a sufficiently high temperature even if the universe started out
in the supersymmetric vacuum initially. 
Note that inclusion of thermal effects from a visible sector was also partly studied in~\cite{Abel:2006my}.
However, an arbitrary high reheating temperature is allowed in the analysis
of these early works and
this may not be
compatible with the cosmological gravitino problem
~\cite{Weinberg:1982zq}.  
In the next section, we will 
elaborate such cosmological evolution of the ISS
model by including thermal corrections from particles in the ISS sector,
a visible sector and partly a messenger sector in light of the cosmological gravitino problem.
\section{Cosmological phase transitions in the ISS model}\label{vacst}
In the early universe, it is natural to expect
that we were in the supersymmetric vacuum (\ref{suvacu}) where
the scalar component of $\Phi$ has a non-zero expectation value.
In \cite{Abel:2006cr,Abel:2006my,Fischler:2006xh,Craig:2006kx} 
it was shown
that if the ISS sector is in thermal equilibrium in the early universe,
then the thermally corrected effective potential $V_{T}(\Phi,\varphi)$ has a global
minimum at the origin of the field space 
at sufficiently high temperature.
Note that here
we parametrize $\Phi_{ij} = \Phi\delta_{ij}$,
$\varphi_{i} = \varphi \bm{1}_{N_{f}}$ so that
the effective potential
is determined by the two parameters. 
The field $\Phi$ can either
tunnel or roll down on the potential, depending on the reheating
temperature, from the supersymmetric vacuum
$(\Phi_0, 0)$ to the  origin
(0,0) of the field space~\cite{Abel:2006cr,Abel:2006my}.  As the temperature of the
universe decreases, a second order phase transition occurs from the
origin to the supersymmetry breaking vacuum (\ref{brevacu}) rather
than to the supersymmetric vacuum
$(\Phi_0, 0)$~\cite{Fischler:2006xh,Craig:2006kx}.

Consider the potential $V_T(\Phi)$ $(\varphi=0)$
which describes the vacuum structure between 
the supersymmetric one (\ref{suvacu}) and the origin of the field space.
From the superpotentials (\ref{supo}) and (\ref{supo2}),
we obtain the potential at zero temperature
\begin{equation}
  V_{0}(\Phi) = |h^{2}\mu^{4}| N_{f}
  \left( \left( \frac{\Phi}{\Phi_{0}} \right)^{\frac{N_{f}-N}{N}}  - 1
  \right)^{2} \ . \label{zerotem}
\end{equation}
In the early universe, there are thermal corrections to the above potential
from particles in equilibrium 
with $\Phi$.
The thermal corrections to the potential (\ref{zerotem}) at temperature
$T$ is given 
by~\cite{Jackiw:1974cv,Weinberg:1974hy}
\begin{equation}
  \Delta V_{T} (\Phi) = \frac{T^{4}}{2\pi^{2}} \sum_{i} \pm n_{i} 
  \int^{\infty}_{0} dp \, p^{2} \ln \left( 1 \mp e^{-\sqrt{p^{2} + m_{i}^{2}(\Phi) / T^{2} }}
  \right) \ ,  \label{thermal}
\end{equation}
where $n_{i}$ denotes degrees of freedom of the $i$-th particle,
$m_{i}(\Phi)$ represents tree level masses of particles, which may
depend on the field expectation value of $\Phi$, and the upper (lower)
signs are for bosonic (fermionic) particles.  Note that both of
contributions from bosons and fermions are negative and those
magnitude are larger if their masses are lighter.  Note also that the
contributions from a $\Phi$-independent  mass
is just a constant shift to the potential.  We will neglect
such contributions since they do not change the shape of the potential.
We mention that although there is also the
Coleman-Weinberg mechanism~\cite{Coleman:1973jx}, which modifies the
potential even at zero temperature, it is exactly zero around the
supersymmetric vacuum and negligible around $\Phi = 0$ compared with
the thermal correction.

In the next subsection, we will review the cosmological gravitino problem and 
find that there
cannot be a
phase transition 
from the supersymmetric vacuum
$(\Phi_0, 0)$ to the origin (0,0) of the field space for
middle and heavy mass regions of gravitino. 
\subsection{Constraints from the gravitino problem}
\label{secgra}
In order to mediate the
supersymmetry breaking of the ISS sector to a visible sector, one needs to
include a messenger sector, which is charged under some gauge group
of a visible sector. 
To be concrete, we will consider the MSSM as a visible sector here after.
In the gauge mediated supersymmetry breaking of
the ISS model, the gravitino would be the lightest supersymmetric
particle among the relic particles. The mass of the gravitino  $ m_{3/2}
= F/\sqrt{3} M_{P}$ is
determined by the supersymmetry breaking scale $F$ and the reduced Planck
mass scale $M_{P}$~\cite{Martin:1997ns}. In our present case, $F =
V_{{\rm meta}}^{1/2}$ and so
\begin{equation}
  m_{3/2} 
    = \frac{\left( N_{f} -N \right)^{1/2} |h \mu^{2}|}{\sqrt{3}
      M_{P}}
    \ . \label{gravitino}
\end{equation}
Since its mass is suppressed by the Planck scale, the gravitino
usually becomes the lightest supersymmetric particle and its
presence could lead to serious cosmological problems. In order to avoid the
overproduction of gravitinos,
the reheating temperature $T_{{\rm R}}$
must be sufficiently low \cite{Weinberg:1982zq,Hook:2018sai}.
The observation of the cosmic microwave background lensing, the cosmic
shear~\cite{Osato:2016ixc}, the Lyman-$\alpha$~\cite{Viel:2005qj}, and
the light-element photodestruction~\cite{Kawasaki:2008qe} have
excluded certain mass regions of the gravitino for any reheating
temperature.  Remaining mass regions are $m_{3/2} < 4.7\,{\rm eV}$, 
$0.4 \,{\rm keV} \lesssim m_{3/2} \lesssim 1 \,{\rm GeV}$, and $ 
600\,{\rm GeV} \lesssim m_{3/2}$~\cite{Hook:2018sai}.  For the
middle region of $0.4 \,{\rm keV} \lesssim m_{3/2} \lesssim 1 \,{\rm GeV}$ 
and the heavy region of $ 600 \,{\rm GeV} \lesssim m_{3/2}$,
higher reheating temperatures compared with the supersymmetry breaking
scales are excluded to avoid the overproduction of gravitinos~\cite{Moroi:1993mb},
and $T_{{\rm R}}$ is at most upper bounded as~\cite{Hook:2018sai}
\begin{equation}
  T_{{\rm R}} < 0.1 \times V_{{\rm meta}}^{1/4} \  .  \label{gracon}
\end{equation}
On the other hand, we do not have any upper bound on the reheating
temperature in the light gravitino mass region 
$m_{3/2} < 4.7 \,{\rm eV}$.

The upper bound \eq{gracon} on the reheating temperature has important
implications on the thermal phase transitions.
The thermal correction \eq{thermal}
to the potential is controlled by $T^4$ times a sum of integral factors.
The bosonic and fermionic integrals vanish at $T=0$ and reach their respective
maximum values of order 1 at large $T$.
Previously in \cite{Abel:2006cr,Abel:2006my}, it was assumed that
the reheating temperature can be as high as one wishes, and as a result
the correction \eq{thermal} is strong  enough to drive the phase transition
from the supersymmetric vacuum to the metastable vacuum. However,
this assumption should be reexamined in view of the gravitino constraint
\eq{gracon} on the reheating temperature.
In fact for temperature below the reheating temperature \eq{gracon},
the thermal correction is  negligible compared to the
zero temperature potential
as long as we do not have unnaturally large number of species $\sum_{i} \pm n_{i} \gg 10^{4}$.
Therefore, we conclude that in the middle and heavy
region of gravitino mass,
the supersymmetric vacuum will remain as the global minimum of the theory
and the desired phase transition
cannot occur.
As a result, we will
concentrate in the rest of the paper on the light gravitino mass region
\be \label{light-mass}
m_{3/2} <4.7 {\rm eV} \ ,
\ee
where there is no upper bound on the reheating temperature.

Finally, let us comment that although there could be massless particles as
Nambu-Goldstone bosons in the ISS sector, which we neglect in the thermal
correction (\ref{thermal}),
they are harmless. They can be
consistent with the upper limits on the abundance of radiation
components~\cite{Cyburt:2015mya} if they decoupled much before the epoch
of the big bang nucleosynthesis ($\sim 0.1$\,MeV). This would be always
possible by tuning parameters in the messenger sector.
\subsection{Evolution of finite temperature effective potential}
\label{secT}
The thermal potential \eq{thermal} is obtained by including the 
contributions  of all particles whose masses
depend on $\Phi$ in the ISS sector,
the MSSM sector and a messenger sector.

Let us first list the
particles in the ISS sector whose masses have non-trivial
dependence on $\Phi$. 
The discussion basically follows the
earlier works of \cite{Abel:2006cr,Abel:2006my}, but is different in some
details.  The $N_{f}$ flavors of $\varphi$ and $\tilde{\varphi}$ are
massless at $\Phi=0$, but they obtain masses of
\begin{equation}
  m_{\varphi} =  h\Phi \ ,  \label{pvarphi}
\end{equation}
when $\Phi$ has a nonzero expectation value.
In the $SU(N)$ gauge sector,
the gauginos and the gauge fields get masses from the condensate
(\ref{conden}) and we have
\begin{equation}
  m_{{\rm gauge}}
                  = \frac{\left( 32 \pi^{2} \right)^{1/3}}{2 (N^{2}-1)} 
                  \left( \frac{h^{N_{f}} {\rm det}\Phi}{\Lambda_{m}^{N_{f}-3N}}
                  \right)^{1/3N}
            = \frac{\left( 32 \pi^{2} \right)^{1/3}}{2 (N^{2}-1)}  
            \epsilon^{(N_{f}-3N)/3N}  \left(\frac{m_{\varphi}}{\mu}\right)^{N_{f}/3N}
            \mu \ .  
       \label{gauge}
\end{equation}
Because the gauge confinement occurs when $\varphi$ fields can be
integrated out, 
we require the condition $ m_{{\rm gauge}} < m_{\varphi}$.
Note that the condition is necessary for the existence of
the supersymmetric vacuum (\ref{suvacu}).
Substituting Eqs.\,(\ref{pvarphi}) and (\ref{gauge}) into Eq.\,(\ref{thermal}),
we get the thermal correction from the ISS sector to the finite temperature
effective potential.
Then the degrees of freedom are evaluated as $\sum_{i} \pm n_{i} = \pm 4NN_{f}$
for $\varphi$ and $\tilde{\varphi}$,
and $\sum_{i} \pm n_{i} = \pm 2(N^{2} - 1)$ for the gauge sector respectively.

Next, we consider a messenger sector
to mediate the supersymmetry breaking to the MSSM sector.
It will turn out that the imposition of the light gravitino mass constraint \eq{light-mass}  
places constraints on the construction of a messenger sector.
Let us illustrate this with
the simplest model of the gauge mediation for the ISS model
\cite{Murayama:2006yf}.
Consider messenger superfields $f$ and $\tilde{f}$ which are charged
under a gauge group of a visible sector.
The messenger fields interact with the ISS sector through a superpotential
\begin{equation}
  W_{{\rm mess}} = \left( M + \lambda {\rm Tr}\Phi  \right) f \tilde{f} \ ,
  \label{supomess}
\end{equation}
where $M$, $\lambda$ are constants and other possible higher order terms have
been omitted. The soft supersymmetry breaking term
is given by the ratio of the $F$-term to the scalar expectation value in 
the superpotential (\ref{supomess}) at the metastable vacuum
(\ref{brevacu})~\cite{Martin:1997ns}.
In our case, it is
\begin{equation}
  m_{{\rm soft}} \sim 
     \frac{g^2}{16\pi^{2}} 
      \frac{\lambda V_{{\rm meta}}^{1/2}}{M} \ , \label{soft}
\end{equation}
where $g$ represents generic standard model coupling constants.
In order to avoid the tachyonic instability 
of the scalar components of the messenger fields in the metastable vacuum, 
the messenger mass scale $M$ is required to satisfy the condition
\cite{Murayama:2006yf}
\begin{equation}
  M > \lambda^{1/2} V_{{\rm meta}}^{1/4} \ .
\end{equation}
Using Eqs.\,(\ref{gravitino}) and (\ref{soft}),
the above inequality can be rewritten to give a lower bound on the gravitino
mass:
\begin{equation}
  m_{3/2} > \frac{(16\pi^{2}/g^2)^{2}}{\sqrt{3}\lambda}
  \frac{m_{{\rm soft}}^{2}}{M_{{P}}} \ . \label{grainq}
\end{equation}
It should be mentioned that from the view point of the electric dual of
the ISS model, 
$\lambda$ naturally takes a small value because it is proportional to a
ratio of
two high energy scales like
$\lambda \propto \Lambda_{{\rm m}}/M_{{P}}$~\cite{Murayama:2006yf}.
At most, one would like to take $\lambda\lesssim 1$ in order
to ensure perturbative treatment of Eq.\,(\ref{supomess}).
Since we have never observed superpartner
particles in particle colliders such as the LHC,
let us fix the soft mass to be $m_{{\rm soft}} \gtrsim  3\,{\rm TeV}$, 
which is slightly above the current lower limits~\cite{Baer:2020kwz}.
Taking also $g\lesssim 1$ in order to maintain the validity of perturbation
theory, then the inequality (\ref{grainq}) reads
\begin{equation}
  m_{3/2} \gtrsim 50 \ {\rm eV} \ .  \label{grainq2}
\end{equation}
The condition excludes the light gravitino window, $m_{3/2} < 4.7$\,eV,
in the ISS model.
On the other hand, as we have discussed in the section \ref{secgra}
for the middle gravitino mass range 
$0.4 \,{\rm keV} \lesssim m_{3/2} \lesssim 1 \,{\rm GeV}$ and the heavy
gravitino mass range $ 600 \,{\rm GeV} \lesssim m_{3/2}$,
the thermal effects to the
potential are too small to give rise to a phase transition
from the supersymmetric vacuum
to the origin $\Phi=0$ due to
the reheating temperature condition (\ref{gracon}).  
Therefore, the ISS model with the
simplest gauge mediation scenario~\cite{Murayama:2006yf}
is not suitable for a metastable supersymmetry breaking scenario
if the supersymmetric
vacuum was chosen initially at the early universe.
In general, in order to allow for the desired supersymmetry
breaking phase transition,  
a gauge mediation model,
e.g.~\cite{Yoshimatsu:2019zfv},
which is compatible with the
light mass gravitino region \eq{light-mass} is needed.
This
can always be arranged and the details are
model dependent, see for example \cite{Meade:2008wd,Carpenter:2008wi}.
We remark that the
mass of messenger fields
usually depends on the expectation value
of $\Phi$, e.g. through a superpotential like that of 
Eq.\,(\ref{supomess})~\cite{Martin:1997ns,Meade:2008wd,Carpenter:2008wi},
such that it is 
lighter at $\Phi = 0$ than at the supersymmetric vacuum
$\Phi=\Phi_{0}$.  Thus, the
thermal contribution of the messenger sector generally tends to assist
the phase transition from the supersymmetric vacuum to $\Phi = 0$; namely,
neglecting the contributions from the messenger sector will not
spoil our following discussion
on the existence of the desired phase transition.
Therefore in this paper,
in order to investigate the general features
of the ISS model of metastable supersymmetry breaking,
we will be entitled not to specify the
messenger sector and 
focus only on the ISS and the MSSM sectors.

Finally, we consider the MSSM sector.
We expect that the both of the standard
model particles and the superpartners are massless in the
supersymmetric vacuum.  However the superpartners obtain masses when
they are far away from the supersymmetric vacuum along the direction
of $\Phi$.  One would then expect that they tend to stabilize the
supersymmetric vacuum since their masses are lighter at the
supersymmetric vacuum than at the origin $\Phi = 0$.  From the
superpotentials (\ref{supo}) and (\ref{supo2}), we can deduce the $F$-term
of $\Phi$ to be
\begin{equation}
  F_{\Phi}|_{\varphi=\tilde{\varphi}=0} = N_{f}^{1/2} |h\mu^{2}| 
  \left| 1 -  \epsilon^{ (N_{f}-3N)/N}   \left(
  \frac{h\Phi}{\mu} \right)^{(N_{f}-N)/N} \right| \ .
\end{equation}
As a result, the tree level mass of superpartner particles along the
$\Phi$ direction is given by~\cite{Martin:1997ns}
\begin{equation}
  m_{{\rm SP}} \sim 
     \frac{g^{2}}{16\pi^{2}} 
      \frac{\lambda F_{\Phi}|_{\varphi=\tilde{\varphi}=0}}{M}\ 
      \sim \frac{ m_{{\rm soft}}\,
        F_{\Phi}|_{\varphi=\tilde{\varphi}=0} }
           {V_{{\rm meta}}^{1/2}} \ ,  \label{SP}
\end{equation}
where $M$ is the messenger mass scale in Eq.\,\eq{supomess} and
Eq.\,(\ref{soft}) has been used to get the second expression.
From the fact that we have never observed superpartner particles in
particle colliders such as the LHC, let us fix the soft mass  to be
 $m_{{\rm soft}} \sim 3$\,TeV.
For simplicity, we will assume that all the superpartner
particles have the same mass of (\ref{SP}).
Then the thermal corrections to the effective potential is obtained by
summing over the $94$ and $32$ degrees of freedom for the bosonic
and the fermionic particles other than the standard model particles in the MSSM
sector. 
Note that the thermal effect from the standard model Higgs is negligible 
because its mass must be smaller than $m_{{\rm soft}}$.

We are now ready to calculate the thermal corrections (\ref{thermal}) from 
the ISS and the MSSM sectors. 
The finite temperature effective
potential is given by
\begin{equation}
  V_{T}(\Phi) = V_{0}(\Phi) + \Delta V_{T}(\Phi) \ ,
\end{equation}
where
\begin{eqnarray}
   \Delta V_{T} (\Phi) =\frac{T^{4}}{2\pi^{2}} \Bigg[  &\pm& 4NN_{f} 
     \int^{\infty}_{0} dp \, p^{2} \ln \left( 1 \mp e^{-\sqrt{p^{2} +
         m_{\varphi}^{2}(\Phi) / T^{2} }} \right) 
\nonumber     \\   
     &\pm&     2(N^{2} - 1)
     \int^{\infty}_{0} dp \, p^{2} \ln \left( 1 \mp e^{-\sqrt{p^{2} +
         m_{{\rm gauge}}^{2}(\Phi) / T^{2} }} \right) 
  \nonumber \\
     &+& 96
  \int^{\infty}_{0} dp \, p^{2} \ln \left( 1 - e^{-\sqrt{p^{2} +
      m_{{\rm SP}}^{2}(\Phi) / T^{2} }} \right)  \nonumber \\
     &-& 32
     \int^{\infty}_{0} dp \, p^{2} \ln \left( 1 + e^{-\sqrt{p^{2} +
         m_{{\rm SP}}^{2}(\Phi) / T^{2} }} \right)  \Bigg] \ .
\end{eqnarray}
\begin{figure}[ht]
\centering
\includegraphics[width=14.0cm]{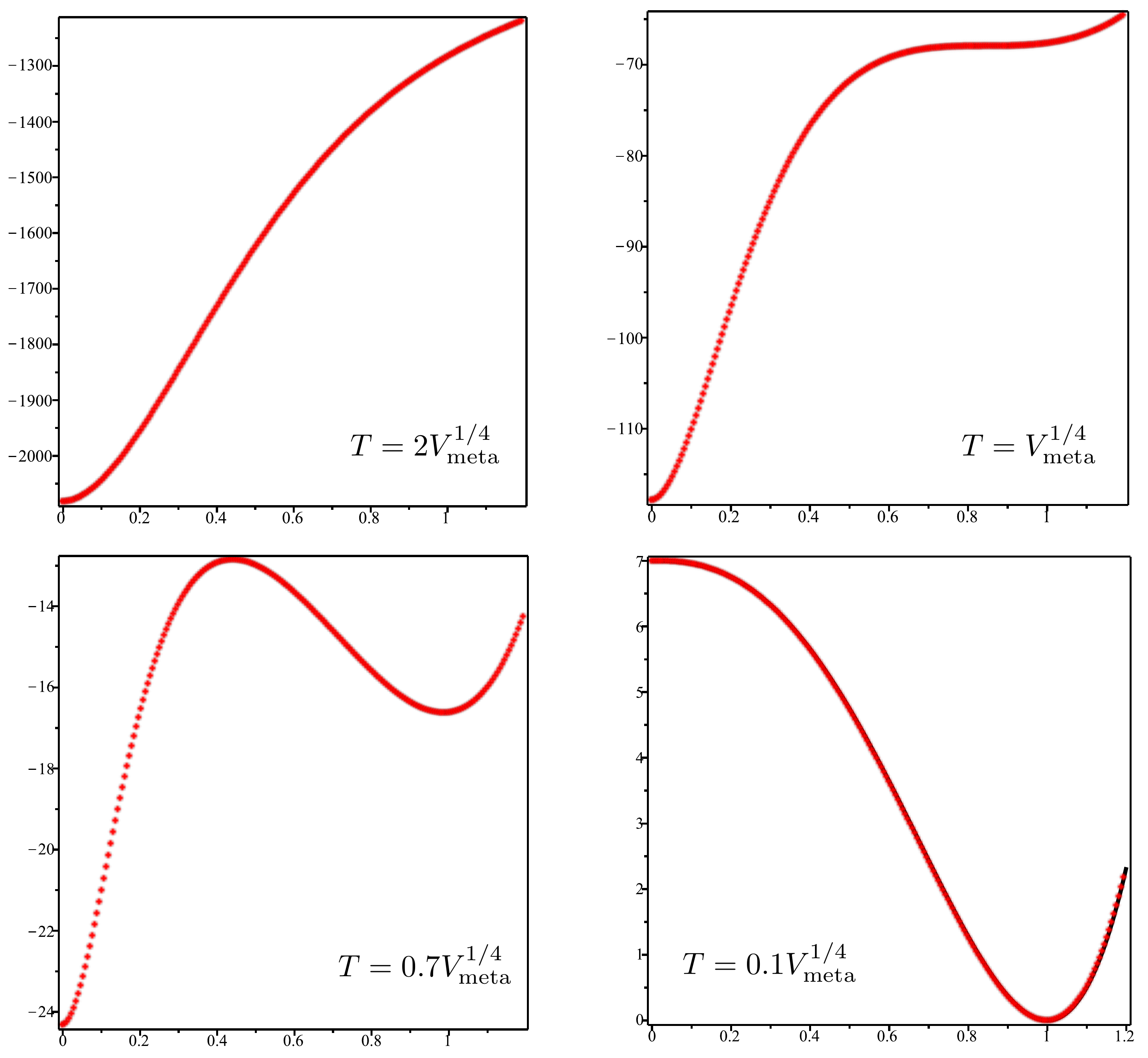}
\caption{Temperature dependence of the effective potential $V_{T}
  (\Phi)$ is shown for the gravitino mass of $m_{3/2} = 1\,{\rm eV}$.
  The horizontal and the longitudinal axes are normalized as
  $\Phi/\Phi_{0}$ and $V_{T}(\Phi)/|h^{2}\mu^{4}|$, respectively.
  Each parameter is set as $h=1,\ \Lambda_{m}=10^{10}\,{\rm
    GeV},\ N=2,\ N_{f}=7$.  When $T > V_{{\rm meta}}^{1/4}$, the
  origin is only a minimum.  Around $T = V_{{\rm meta}}^{1/4}$, an
  inflection point comes along, namely a second minimum around the
  supersymmetric vacuum appears.  When the temperature is right below
  the supersymmetry breaking scale $V_{{\rm meta}}^{1/4}$, we have two
  local minima.  In the figure of $T = 0.1 V_{{\rm meta}}^{1/4}$, the
  zero temperature potential $V_{0}(\Phi)$ is also depicted by a
  black line for reference.  It shows that the thermal correction is
  already negligible at $T = 0.1 V_{{\rm meta}}^{1/4}$.  }
\label{potential}
\end{figure}
We are interested in
the temperature dependence of the finite temperature effective
potential in the light gravitino mass region.
In Fig.\,\ref{potential}, we show the
temperature dependence of the potential for 
the gravitino mass $m_{3/2} = 1\,{\rm eV}$.
The corresponding supersymmetry breaking scale is
$V_{{\rm meta}}^{1/4} = 6.5 \times 10^{4}$\,GeV and
the parameters of the ISS model are taken as
$h=1,\  \Lambda_{m}=10^{10}$\,GeV,\  $N=2,\  N_{f}=7$.
From Fig.\,\ref{potential}, we can see that the origin is only a
minimum of the potential
when the temperature is high compared with the 
supersymmetry breaking scale.  During this epoch, the scalar field can
roll down to the origin.  Around $T=V_{{\rm meta}}^{1/4}$, an
inflection point appears around the supersymmetric vacuum.  We will
call this temperature $T_{*}$.  Below $T_{*}$, there is a critical
temperature $T_{{\rm crit}}$ at which the energy of the two vacua
degenerates.  For $T_{{\rm crit}} < T < T_{*}$, tunneling process can
occur from around the supersymmetric vacuum to the origin.  For $T <
T_{{\rm crit}}$, the supersymmetric vacuum becomes the global minimum and 
we can no longer have any more phase transitions from the supersymmetric vacuum
to the origin.  Instead, one may have tunneling from the
origin to the supersymmetric vacuum.  However, it is known that the
tunneling would not occur because the phase transition from the origin
to the metastable vacuum (\ref{brevacu}) is more
efficient~\cite{Fischler:2006xh,Craig:2006kx}.
If the reheating
temperature $T_{{\rm R}}$ is lower than $T_{{\rm crit}}$, the universe will
stay in the supersymmetry vacuum and  will never 
come to the metastable vacuum.
A phenomenologically interesting case is when
\be
T_{{\rm crit}} < T_{{\rm R}} < T_{*} \ ,
\ee
where many bubbles can be nucleated from the 
tunneling process and abundant gravitational waves could be produced.
We will study this possibility in the next section.
\section{Gravitational waves from bubble nucleation}\label{GW}
As shown in the previous section, there could be tunneling process from
the false vacuum around 
$\Phi=\Phi_{0}$ to the true vacuum at $\Phi=0$.
Indeed it happens if the reheating temperature is in the region,
$T_{{\rm crit}} < T_{{\rm R}} < T_{*}$, and then
bubbles are nucleated%
\footnote{We assume an instantaneous reheating in this paper so that
  the reheating completes in shorter duration compared with other
  physical scales like the Hubble constant.}.  In this section, we
calculate gravitational waves produced by the dynamics of
bubbles~\cite{Turner:1990rc,Hogan:1986qda,Kosowsky:1992rz,
  Kamionkowski:1993fg,Turner:1992tz}.
For simplicity, we will calculate gravitational wave production at
$T_{\rm R} = T_{*}$ at which the gravitational wave production is
expected to be the most efficient.

\begin{figure}[ht]
\centering
\includegraphics[width=8cm]{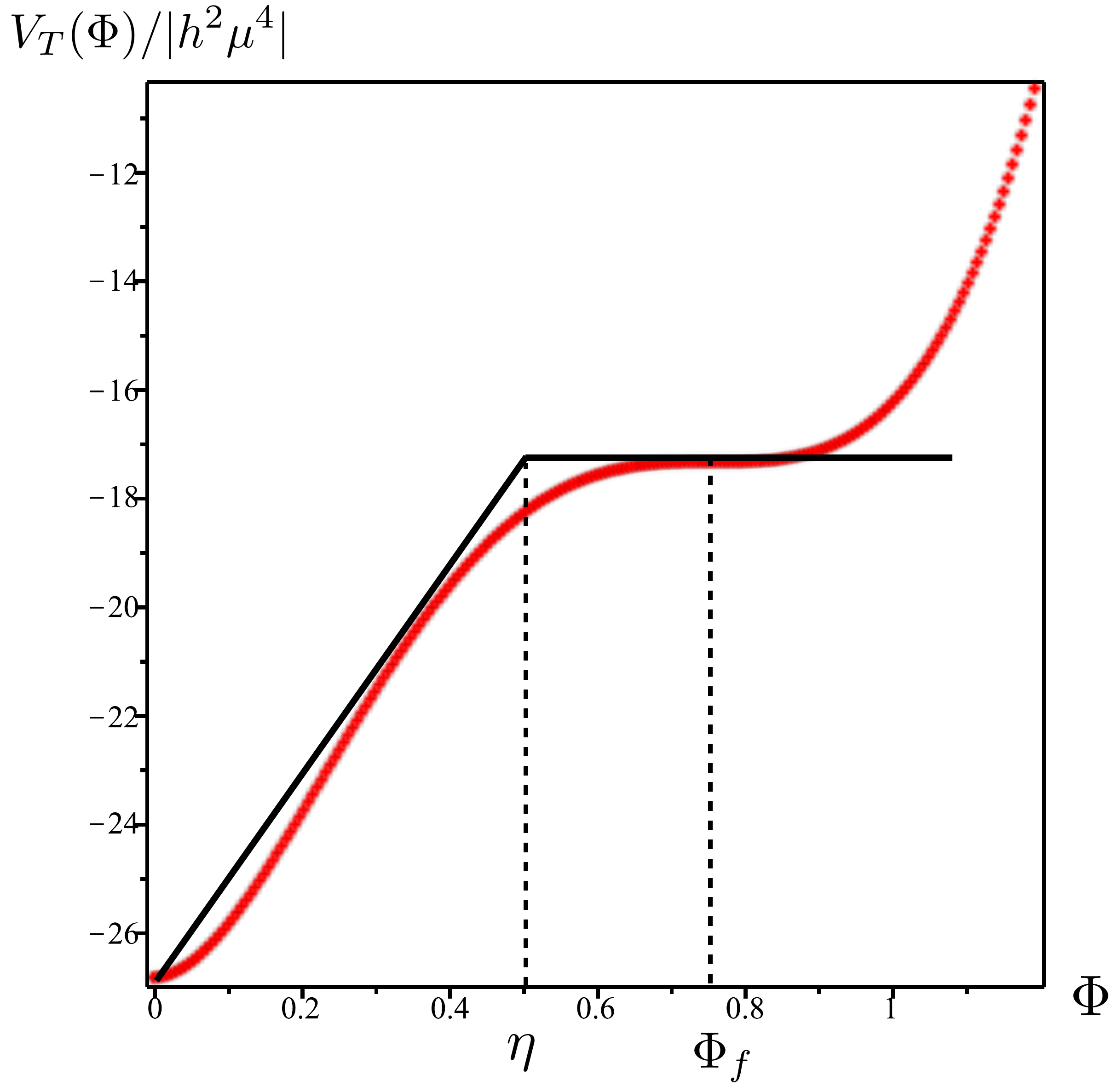}
\caption{The red dot line represents the finite temperature effective potential
for a parameter set of
$m_{3/2} = 1\,{\rm eV},\  h=1,\  \Lambda_{m}=10^{10}\,{\rm GeV},\  N=5,\
N_{f}=16$.
The horizontal axis is normalized by $\Phi_{0}$.
The black line represents the flat potential approximation (\ref{pote})
with $\eta=2\Phi_{f}/3 $.
}
\label{flat}
\end{figure}

The transition rate per unit volume in thermal universe is given 
by~\cite{Linde:1981zj}
\begin{equation}
   \Gamma \simeq e^{-S_{3}/T} \ ,  \label{rate}
\end{equation}
where $S_{3}$ is a three dimensional bounce action for 
$O(3)$ symmetric configurations, which gives  
the lowest action~\cite{Coleman:1977th,Linde:1981zj} and dominates the transition.
The classical equation for a bounce solution is given by
\begin{equation}
  \Phi''(r) + \frac{2}{r} \Phi'(r) = \frac{1}{N_{f}} \partial_{\Phi} V_T(\Phi) \ ,
  \label{boun}
\end{equation}
where $r$ is the
spatial radial distance.
Since it is difficult to obtain an analytic form of the bounce solution for
general potentials,
we approximately parametrize our potential as~\cite{Lee:1985uv,Abel:2006cr}:
\begin{equation}
  V_T (\Phi) =  N_{f} K \, 
   (\Phi - \eta)  \theta(\eta - \Phi) \ ,  \label{pote}
\end{equation}
where $K$ and $\eta$ are positive 
constants, $\theta$ represents the step
function and $N_{f}$ has been put for later convenience.  For the
potential (\ref{pote}), one can find an analytic bounce solution of
the equation (\ref{boun})
with the
bounce
boundary condition $\Phi(\infty) = \Phi_f$ and
$\frac{d\Phi(r)}{dr}|_{r=0} = 0$.
Here $\Phi_f$ denotes the location of
the supersymmetric vacuum after the
finite temperature thermal effects are included.
The bounce
solution is given by 
\begin{equation}
  \Phi(r)=
  \begin{cases}
    \frac{K}{6} (r^2 -r_m^2) +\eta
    & \text{for \ $r\leq r_m$\ ,} \\
    \Phi_f - \frac{K r_m^3}{3 r} 
    & \text{for \  $r \geq r_m$\ ,} 
      \end{cases}
\end{equation}
where $r_m = \sqrt{3(\Phi_f - \eta)/K}$. This bounce solution is parametrized
by the position of the false vacuum $\Phi_f$, the turning point $\eta$
of the potential, and $K$ which characterizes the tilt of the potential.
We note that
for a wide range of parameters, which is compatible with the
light gravitino mass region and of interest to us for the observation
of gravitational waves,
the position of the false vacuum
can be approximated by $\Phi_f = 3 \eta/2$ (see Fig.\,\ref{flat})
to a very good approximation.
Then the bounce action is given by
\begin{equation}
  \frac{S_{3}}{T} = \frac{\sqrt{6} \pi N_{f} \eta^{5/2}}{5 \sqrt{K} T} \ .
\end{equation}

As a consequence of the phase transition from the false vacuum to the
true vacuum, bubbles are nucleated and gravitational waves can be
produced~\cite{Turner:1990rc,Hogan:1986qda,Kosowsky:1992rz,Kamionkowski:1993fg,
  Turner:1992tz}.
The dynamics of the bubbles and the produced gravitational waves are
characterized
by two parameters. The first one is the
vacuum energy density released in the phase transition to the energy
density of the radiation bath~\cite{Caprini:2015zlo},
\begin{equation}
  \alpha = \frac{V_{f} - V_{T}(\Phi=0)}{g_{*}\pi^{2}T_{*}^{4}/30} \ ,
\end{equation}
where $g_{*}$ is the effective degrees of freedom of particles at the
temperature $T_{*}$. The second parameter is 
\begin{equation}
  \beta = - H_{*}T_{*} \left( \frac{d}{dT} \frac{S_{3}}{T} \right)_{T=T_{*}} \ ,
\end{equation}
which approximately represents the duration of the phase transition~\cite{Turner:1992tz}.
Here $H_{*} = \sqrt{\frac{\pi^{2} g_{*} T_{*}^{4}}{90 M_{P}^{2}}}$ 
is the Hubble constant at the time of phase transition.
We note the sign of the function inside the parenthesis is negative in
the situations we consider, so that $\beta$ has been defined to be
positive.    There are two kinds
of scenarios for the dynamics of nucleated bubbles in the radiation
dominated era~\cite{Caprini:2015zlo}: the runaway bubbles and
the no runaway bubbles.  In the former scenario, the bubble
wall accelerates continuously and approaches the speed of
light.  In the no runaway scenario, the speed of the bubble walls converges to
a value smaller than the speed of light.
A useful
parameter which can be used to discriminate between
the two scenarios is~\cite{Espinosa:2010hh}
\begin{equation}
  \alpha_{\infty} := \frac{30}{24\pi} \frac{\sum_{i} c_{i}
    \Delta m_{i}^{2} }{g_{*}T_{*}^{2}} \  ,
\end{equation}
where the sum is taken for particles which is lighter in the false
vacuum than in the true vacuum, $c_{i}$ stands for degrees of freedom
of the particle (a factor of $1/2$ is additionally multiplied for
fermions) and $\Delta m_{i}$ is
the difference in mass of the particle
between the two vacua.
If $\alpha > \alpha_{\infty}$, the energy of the phase transition can
convert to the acceleration of bubble walls and the runaway bubbles
scenario is realized.  If $\alpha < \alpha_{\infty}$, all the phase
transition energy is deposited into the plasma and the no runaway
bubbles scenario is realized.  
In our case, the superpartner particles in
the MSSM sector contribute to $\alpha_{\infty}$.
We will only consider the case in which $m_{{\rm SP}} \sim {\rm TeV} \ll T_{*}$.  
This implies that $\alpha \gg
\alpha_{\infty}$ and so we have the runaway bubbles scenario.
In this case, 
bubble collisions~\cite{Kosowsky:1991ua,Kosowsky:1992rz,
  Kosowsky:1992vn,Kamionkowski:1993fg,Caprini:2007xq,Huber:2008hg}
would be the dominant source of gravitational waves compared with the
contributions from the sound 
wave~\cite{Hindmarsh:2013xza,Giblin:2013kea, Giblin:2014qia,Hindmarsh:2015qta}
and the magnetohydrodynamic 
turbulence~\cite{Caprini:2006jb,Kahniashvili:2008pf,Kahniashvili:2008pe,
  Kahniashvili:2009mf,Caprini:2009yp}  in the plasma.
For bubble collisions, the gravitational wave spectrum in terms of the energy density 
is given by~\cite{Huber:2008hg,Caprini:2015zlo}
\begin{equation}
  h_{0}^{2}\Omega_{f}(f) = 1.67 \times 10^{-5}
                        \left( \frac{H_{*}}{\beta} \right)^{2}
                        \left( \frac{\kappa \alpha}{1 + \alpha} \right)^{2}
                        \left( \frac{100}{g_{*}} \right)^{1/3}
                        \left( \frac{0.11 v_{w}^{3}}{0.42+v_{w}^{2}} \right)
                        \left( \frac{3.8 (f/f_{p})^{2.8}}{1 + 2.8 (f/f_{p})^{3.8}}
                        \right)\ ,  \label{fit}
\end{equation}
where $\kappa = 1 - \alpha_{\infty}/\alpha$ and
we will assume the bubble wall velocity $v_{w}$ is the speed of light from now on.
$f_{p}$ is the peak frequency of the gravitational wave spectrum at present,
\begin{equation}
  f_{p} = 16.5 \times 10^{-6} \, {\rm Hz} 
           \left( \frac{0.62}{1.8 - 0.1 v_{w} + v_{w}^{2}} \right)
           \left( \frac{\beta}{H_{*}} \right)
           \left( \frac{T_{*}}{100\, {\rm GeV}} \right)  \label{peak}
           \left( \frac{g_{*}}{100} \right)^{1/6} \ .
\end{equation}
\begin{figure}[ht]
\centering
\includegraphics[width=16.0cm]{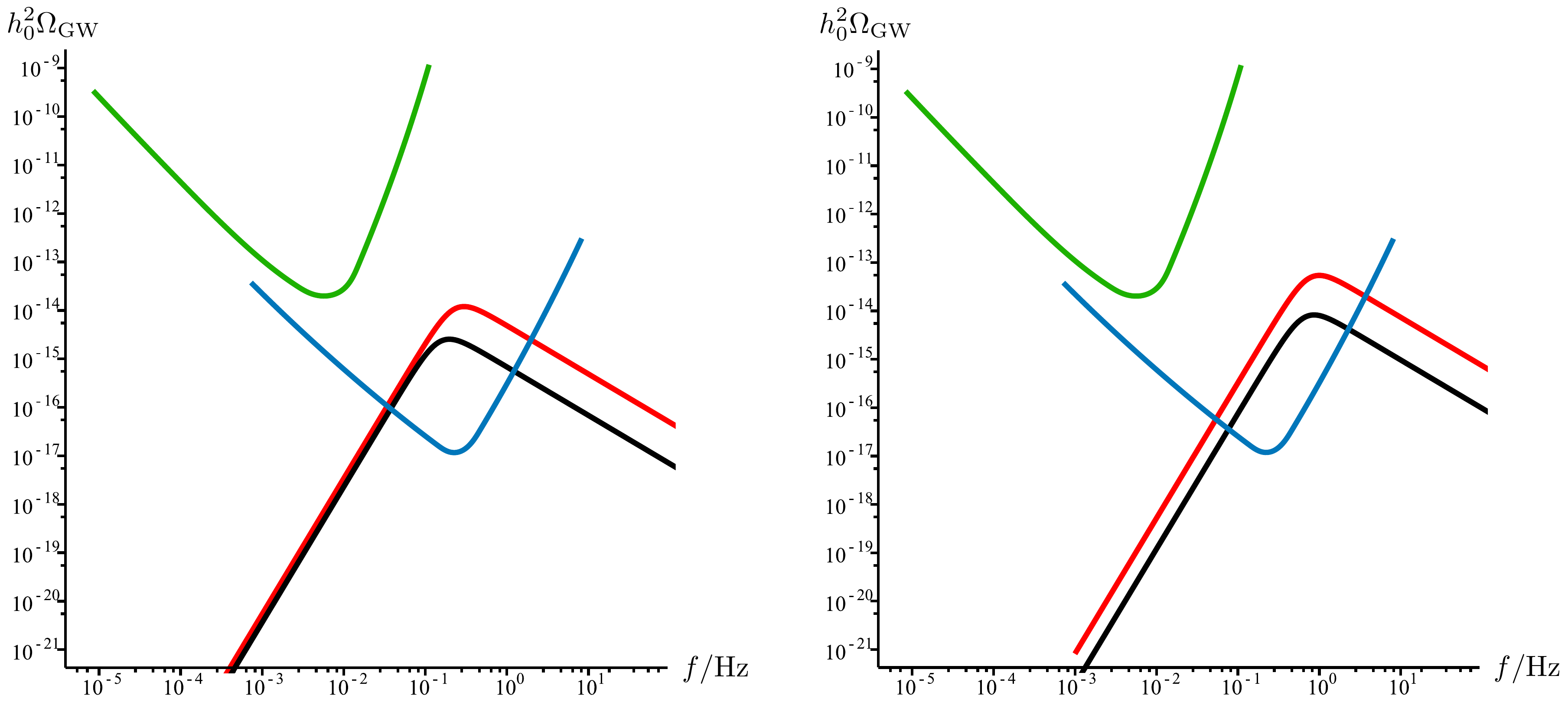}
\caption{Gravitational wave spectra from bubble collisions for $h = 1,\
  \Lambda_{m}=10^{10}$\,GeV are depicted.  The gravitino masses are set
  $m_{3/2} = 0.01$\,eV in
  the left side figure and $m_{3/2} = 1$\,eV in the right side
  figure. In each figure, the red and black lines represent the
  gravitational wave spectra for $N=2, N_{f}=7$ and $N=5, N_{f}=16$,
  respectively.  The green line is the sensitivity curve of LISA of
  the C1 configuration~\cite{Caprini:2015zlo}.  The blue line is the
  sensitivity curve of DECIGO for two
  clusters~\cite{Kawamura:2020pcg}.
}
\label{GWh1}
\end{figure}
\begin{figure}[ht]
\centering
\includegraphics[width=16.0cm]{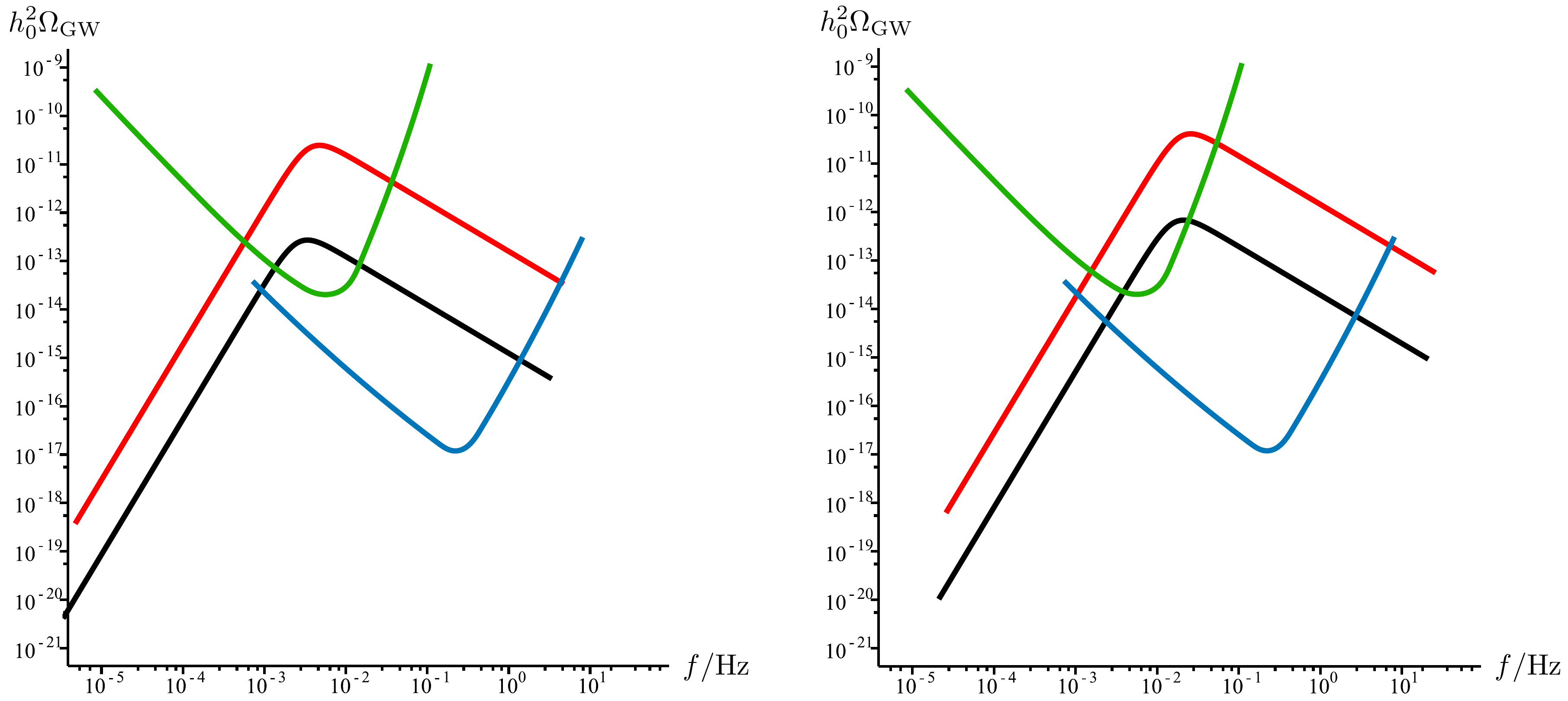}
\caption{Gravitational wave spectra from bubble collisions for $h = 3,\
  \Lambda_{m}=10^{10}$\,GeV are depicted.
  The gravitino masses are set $m_{3/2} = 0.01$\,eV in the left side figure and
  $m_{3/2} = 1$\,eV
  in the right side figure. In each figure,
  the red and black lines represent the gravitational wave spectra
for $N=2, N_{f}=7$ and $N=5, N_{f}=16$, respectively.
The green line is the sensitivity curve of LISA of the C1
configuration~\cite{Caprini:2015zlo}.
The blue line is the sensitivity curve of DECIGO for two 
clusters~\cite{Kawamura:2020pcg}.
}
\label{GWh3}
\end{figure}
We calculated several gravitational wave spectra from bubble
collisions for the gravitino mass $m_{3/2} =0.01$\,eV ($V_{{\rm
    meta}}^{1/4} = 6.5 \times 10^{3}$\,GeV) and $1$\,eV ($V_{{\rm
    meta}}^{1/4} = 6.5 \times 10^{4}$\,GeV) with the fitting formula
(\ref{fit}) and (\ref{peak}).  Then $\alpha$ and $\beta$ were
numerically evaluated from potentials at $T=T_{*}$.  The results are
depicted in Figs.\,\ref{GWh1} and \ref{GWh3}.  In Fig.\,\ref{GWh1},
the energy density of gravitational waves is depicted for
$h=1,\ \Lambda_{m}=10^{10}$\,GeV.  The gravitino masses are set $m_{3/2} =
0.01$\,eV in the left side figure and $m_{3/2} = 1$\,eV in the right side
figure.  The red line and the black line correspond to the parameter
sets of $N=2, N_{f}=7$ and $N=5, N_{f}=16$, respectively.  One can see
that both of the produced gravitational waves are within the sensitivity
of DECIGO~\cite{Kawamura:2020pcg}, whose sensitivity is represented by
the blue line.  In Fig.\,\ref{GWh3}, the energy density of
gravitational waves is depicted for $h=3,\ \Lambda_{m}=10^{10}$\,GeV.  The
gravitino masses are set $m_{3/2} = 0.01$\,eV in the left side figure
and $m_{3/2} = 1$\,eV
in the right side figure.
Again, the red line and the black line correspond to the parameter
sets of $N=2, N_{f}=7$ and $N=5, N_{f}=16$, respectively.  The spectra
are within the sensitivity of DECIGO~\cite{Kawamura:2020pcg} (blue
line) and LISA~\cite{Caprini:2015zlo} (green line).  From
Figs.\,\ref{GWh1} and \ref{GWh3}, one can see that the produced
gravitational waves for the stated parameter sets are detectable with the
future gravitational wave interferometers.  One finds that amplitude
of gravitational waves becomes larger for bigger $h$.  We also see
that because lighter gravitino masses correspond to lower
supersymmetry breaking energy scales, the peak frequency of the
spectra tend to be lower for lighter gravitino masses.  
Potentially, we could explore the parameters of the ISS model of the
metastable supersymmetry breaking scenario with gravitational wave
observations.
We finally mention that since adding contributions from a messenger sector 
to the thermal correction (\ref{thermal}) assists the phase transitions as explained in
the section \ref{secT},
more gravitational wave production is expected if a messenger sector is considered explicitly.
\section{Conclusion}
In this paper, we first elaborated the cosmological evolution of the vacuum structure of the ISS model 
in light of the cosmological gravitino problem.
The incorporation of the gravitino constraints is crucial to our analysis. 
We showed that in the middle gravitino
mass region $0.4 \,{\rm keV} \lesssim m_{3/2} \lesssim 1 \,{\rm GeV}$
and the heavy gravitino mass region $ 600 \,{\rm GeV} \lesssim
m_{3/2}$,
there are
insufficient thermal effects to bring forth the desired
phase transitions from the supersymmetric vacuum to the metastable vacuum
since the reheating temperature is stringently bounded.
Thus only in the light gravitino mass region,
$m_{3/2} < 4.7 \,{\rm eV}$, we have enough thermal
effects to allow for the phase transition. 
This is achieved by either rolling down potential or tunneling processes depending on 
the reheating temperature.

We also calculated gravitational wave spectra associated with the tunneling 
from the supersymmetry vacuum to the metastable vacuum.
Abundant gravitational waves could be produced by
collisions of runaway bubbles and they are detectable with the future
gravitational wave interferometers like LISA and DECIGO.  This gives
us a unique way to probe the metastable supersymmetry breaking.
Note that although the messenger sector was not included in the calculation
in order to allow a model independent treatment,
our conclusion is robust
because the thermal effects from the messenger sector 
assists the phase transition and more gravitational wave production is
expected in general
when the thermal contributions of the messenger sector are included.

It should be mentioned that we did not consider the mass spectrum of
the two Higgs doublets and superpartner particles seriously, since it
goes beyond the purpose of this paper.  In order to do so, we need to
specify a specific model of the messenger sector for the gauge
mediation.  Fortunately, it seems that there has already been a gauge
mediation model which is compatible with our
scenario~\cite{Yoshimatsu:2019zfv}. 
It would be interesting to study
the correlation between the mass spectrum of superpartner particles
and gravitational wave spectra in detail~\cite{Craig:2020jfv}. 
We note that the employment of a visible sector other than the MSSM 
is straightforward and would not change our conclusion as
long
as the masses of superpartner particles 
are not so heavy compared with the metastable supersymmetry breaking scale.
We also mention that although we calculated
the gravitational wave spectra by
assuming the reheating temperature to be $T_{{\rm R}} = T_{*}$, such a
situation may be necessarily realized if inflation is embedded in the
ISS model~\cite{Craig:2008tv} because $T_{*}$ is usually close to the
supersymmetry breaking scale as shown in Fig.\,\ref{potential}.
It is also interesting to extend our discussion to the ordinal
O'Raifeartaigh type models of metastable supersymmetry
breaking~\cite{Dalianis:2010yk,Dalianis:2010pq}.
We leave these issues for future work.
\begin{acknowledgments}
We thank Kingman Cheung for helpful discussion on the subject.
  C.\,S.\,C.  acknowledge support of this work by
  NCTS and
  the grant 110-2112-M-007-015-MY3 of the
Ministry of Science and Technology of Taiwan.
A.\,I.\ was supported by JSPS KAKENHI Grant Number JP21J00162.
\end{acknowledgments}

\bibliography{ref}

\begin{thebibliography}{74}%
\makeatletter
\providecommand \@ifxundefined [1]{%
 \@ifx{#1\undefined}
}%
\providecommand \@ifnum [1]{%
 \ifnum #1\expandafter \@firstoftwo
 \else \expandafter \@secondoftwo
 \fi
}%
\providecommand \@ifx [1]{%
 \ifx #1\expandafter \@firstoftwo
 \else \expandafter \@secondoftwo
 \fi
}%
\providecommand \natexlab [1]{#1}%
\providecommand \enquote  [1]{``#1''}%
\providecommand \bibnamefont  [1]{#1}%
\providecommand \bibfnamefont [1]{#1}%
\providecommand \citenamefont [1]{#1}%
\providecommand \href@noop [0]{\@secondoftwo}%
\providecommand \href [0]{\begingroup \@sanitize@url \@href}%
\providecommand \@href[1]{\@@startlink{#1}\@@href}%
\providecommand \@@href[1]{\endgroup#1\@@endlink}%
\providecommand \@sanitize@url [0]{\catcode `\\12\catcode `\$12\catcode
  `\&12\catcode `\#12\catcode `\^12\catcode `\_12\catcode `\%12\relax}%
\providecommand \@@startlink[1]{}%
\providecommand \@@endlink[0]{}%
\providecommand \url  [0]{\begingroup\@sanitize@url \@url }%
\providecommand \@url [1]{\endgroup\@href {#1}{\urlprefix }}%
\providecommand \urlprefix  [0]{URL }%
\providecommand \Eprint [0]{\href }%
\providecommand \doibase [0]{http://dx.doi.org/}%
\providecommand \selectlanguage [0]{\@gobble}%
\providecommand \bibinfo  [0]{\@secondoftwo}%
\providecommand \bibfield  [0]{\@secondoftwo}%
\providecommand \translation [1]{[#1]}%
\providecommand \BibitemOpen [0]{}%
\providecommand \bibitemStop [0]{}%
\providecommand \bibitemNoStop [0]{.\EOS\space}%
\providecommand \EOS [0]{\spacefactor3000\relax}%
\providecommand \BibitemShut  [1]{\csname bibitem#1\endcsname}%
\let\auto@bib@innerbib\@empty
\bibitem [{\citenamefont {Grishchuk}(1974)}]{Grishchuk:1974ny}%
  \BibitemOpen
  \bibfield  {author} {\bibinfo {author} {\bibfnamefont {L.~P.}\ \bibnamefont
  {Grishchuk}},\ }\href@noop {} {\bibfield  {journal} {\bibinfo  {journal} {Zh.
  Eksp. Teor. Fiz.}\ }\textbf {\bibinfo {volume} {67}},\ \bibinfo {pages} {825}
  (\bibinfo {year} {1974})}\BibitemShut {NoStop}%
\bibitem [{\citenamefont {Starobinsky}(1979)}]{Starobinsky:1979ty}%
  \BibitemOpen
  \bibfield  {author} {\bibinfo {author} {\bibfnamefont {A.~A.}\ \bibnamefont
  {Starobinsky}},\ }\href@noop {} {\bibfield  {journal} {\bibinfo  {journal}
  {JETP Lett.}\ }\textbf {\bibinfo {volume} {30}},\ \bibinfo {pages} {682}
  (\bibinfo {year} {1979})}\BibitemShut {NoStop}%
\bibitem [{\citenamefont {Ito}\ \emph {et~al.}(2021)\citenamefont {Ito},
  \citenamefont {Soda},\ and\ \citenamefont {Yamaguchi}}]{Ito:2020neq}%
  \BibitemOpen
  \bibfield  {author} {\bibinfo {author} {\bibfnamefont {A.}~\bibnamefont
  {Ito}}, \bibinfo {author} {\bibfnamefont {J.}~\bibnamefont {Soda}}, \ and\
  \bibinfo {author} {\bibfnamefont {M.}~\bibnamefont {Yamaguchi}},\ }\href
  {\doibase 10.1088/1475-7516/2021/03/033} {\bibfield  {journal} {\bibinfo
  {journal} {JCAP}\ }\textbf {\bibinfo {volume} {03}},\ \bibinfo {pages} {033}
  (\bibinfo {year} {2021})},\ \Eprint {http://arxiv.org/abs/2009.03611}
  {arXiv:2009.03611 [astro-ph.CO]} \BibitemShut {NoStop}%
\bibitem [{\citenamefont {Akrami}\ \emph {et~al.}(2020)\citenamefont {Akrami}
  \emph {et~al.}}]{Planck:2018jri}%
  \BibitemOpen
  \bibfield  {author} {\bibinfo {author} {\bibfnamefont {Y.}~\bibnamefont
  {Akrami}} \emph {et~al.} (\bibinfo {collaboration} {Planck}),\ }\href
  {\doibase 10.1051/0004-6361/201833887} {\bibfield  {journal} {\bibinfo
  {journal} {Astron. Astrophys.}\ }\textbf {\bibinfo {volume} {641}},\ \bibinfo
  {pages} {A10} (\bibinfo {year} {2020})},\ \Eprint
  {http://arxiv.org/abs/1807.06211} {arXiv:1807.06211 [astro-ph.CO]}
  \BibitemShut {NoStop}%
\bibitem [{\citenamefont {Ade}\ \emph {et~al.}(2021)\citenamefont {Ade} \emph
  {et~al.}}]{BICEP:2021xfz}%
  \BibitemOpen
  \bibfield  {author} {\bibinfo {author} {\bibfnamefont {P.~A.~R.}\
  \bibnamefont {Ade}} \emph {et~al.} (\bibinfo {collaboration} {BICEP, Keck}),\
  }\href {\doibase 10.1103/PhysRevLett.127.151301} {\bibfield  {journal}
  {\bibinfo  {journal} {Phys. Rev. Lett.}\ }\textbf {\bibinfo {volume} {127}},\
  \bibinfo {pages} {151301} (\bibinfo {year} {2021})},\ \Eprint
  {http://arxiv.org/abs/2110.00483} {arXiv:2110.00483 [astro-ph.CO]}
  \BibitemShut {NoStop}%
\bibitem [{\citenamefont {Arzoumanian}\ \emph {et~al.}(2020)\citenamefont
  {Arzoumanian} \emph {et~al.}}]{NANOGrav:2020bcs}%
  \BibitemOpen
  \bibfield  {author} {\bibinfo {author} {\bibfnamefont {Z.}~\bibnamefont
  {Arzoumanian}} \emph {et~al.} (\bibinfo {collaboration} {NANOGrav}),\ }\href
  {\doibase 10.3847/2041-8213/abd401} {\bibfield  {journal} {\bibinfo
  {journal} {Astrophys. J. Lett.}\ }\textbf {\bibinfo {volume} {905}},\
  \bibinfo {pages} {L34} (\bibinfo {year} {2020})},\ \Eprint
  {http://arxiv.org/abs/2009.04496} {arXiv:2009.04496 [astro-ph.HE]}
  \BibitemShut {NoStop}%
\bibitem [{\citenamefont {Goncharov}\ \emph {et~al.}(2021)\citenamefont
  {Goncharov} \emph {et~al.}}]{Goncharov:2021oub}%
  \BibitemOpen
  \bibfield  {author} {\bibinfo {author} {\bibfnamefont {B.}~\bibnamefont
  {Goncharov}} \emph {et~al.},\ }\href {\doibase 10.3847/2041-8213/ac17f4} {\
  (\bibinfo {year} {2021}),\ 10.3847/2041-8213/ac17f4},\ \Eprint
  {http://arxiv.org/abs/2107.12112} {arXiv:2107.12112 [astro-ph.HE]}
  \BibitemShut {NoStop}%
\bibitem [{\citenamefont {Chen}\ \emph
  {et~al.}(2021{\natexlab{a}})\citenamefont {Chen} \emph
  {et~al.}}]{Chen:2021rqp}%
  \BibitemOpen
  \bibfield  {author} {\bibinfo {author} {\bibfnamefont {S.}~\bibnamefont
  {Chen}} \emph {et~al.},\ }\href {\doibase 10.1093/mnras/stab2833} {\bibfield
  {journal} {\bibinfo  {journal} {Mon. Not. Roy. Astron. Soc.}\ }\textbf
  {\bibinfo {volume} {508}},\ \bibinfo {pages} {4970} (\bibinfo {year}
  {2021}{\natexlab{a}})},\ \Eprint {http://arxiv.org/abs/2110.13184}
  {arXiv:2110.13184 [astro-ph.HE]} \BibitemShut {NoStop}%
\bibitem [{\citenamefont {Chen}\ \emph
  {et~al.}(2021{\natexlab{b}})\citenamefont {Chen}, \citenamefont {Wu},\ and\
  \citenamefont {Huang}}]{Chen:2021ncc}%
  \BibitemOpen
  \bibfield  {author} {\bibinfo {author} {\bibfnamefont {Z.-C.}\ \bibnamefont
  {Chen}}, \bibinfo {author} {\bibfnamefont {Y.-M.}\ \bibnamefont {Wu}}, \ and\
  \bibinfo {author} {\bibfnamefont {Q.-G.}\ \bibnamefont {Huang}},\ }\href@noop
  {} {\  (\bibinfo {year} {2021}{\natexlab{b}})},\ \Eprint
  {http://arxiv.org/abs/2109.00296} {arXiv:2109.00296 [astro-ph.CO]}
  \BibitemShut {NoStop}%
\bibitem [{\citenamefont {Neronov}\ \emph {et~al.}(2021)\citenamefont
  {Neronov}, \citenamefont {Roper~Pol}, \citenamefont {Caprini},\ and\
  \citenamefont {Semikoz}}]{Neronov:2020qrl}%
  \BibitemOpen
  \bibfield  {author} {\bibinfo {author} {\bibfnamefont {A.}~\bibnamefont
  {Neronov}}, \bibinfo {author} {\bibfnamefont {A.}~\bibnamefont {Roper~Pol}},
  \bibinfo {author} {\bibfnamefont {C.}~\bibnamefont {Caprini}}, \ and\
  \bibinfo {author} {\bibfnamefont {D.}~\bibnamefont {Semikoz}},\ }\href
  {\doibase 10.1103/PhysRevD.103.L041302} {\bibfield  {journal} {\bibinfo
  {journal} {Phys. Rev. D}\ }\textbf {\bibinfo {volume} {103}},\ \bibinfo
  {pages} {041302} (\bibinfo {year} {2021})},\ \Eprint
  {http://arxiv.org/abs/2009.14174} {arXiv:2009.14174 [astro-ph.CO]}
  \BibitemShut {NoStop}%
\bibitem [{\citenamefont {Li}\ \emph {et~al.}(2021)\citenamefont {Li},
  \citenamefont {Shao}, \citenamefont {Wu},\ and\ \citenamefont
  {Yu}}]{Li:2021qer}%
  \BibitemOpen
  \bibfield  {author} {\bibinfo {author} {\bibfnamefont {S.-L.}\ \bibnamefont
  {Li}}, \bibinfo {author} {\bibfnamefont {L.}~\bibnamefont {Shao}}, \bibinfo
  {author} {\bibfnamefont {P.}~\bibnamefont {Wu}}, \ and\ \bibinfo {author}
  {\bibfnamefont {H.}~\bibnamefont {Yu}},\ }\href {\doibase
  10.1103/PhysRevD.104.043510} {\bibfield  {journal} {\bibinfo  {journal}
  {Phys. Rev. D}\ }\textbf {\bibinfo {volume} {104}},\ \bibinfo {pages}
  {043510} (\bibinfo {year} {2021})},\ \Eprint
  {http://arxiv.org/abs/2101.08012} {arXiv:2101.08012 [astro-ph.CO]}
  \BibitemShut {NoStop}%
\bibitem [{\citenamefont {Apreda}\ \emph {et~al.}(2002)\citenamefont {Apreda},
  \citenamefont {Maggiore}, \citenamefont {Nicolis},\ and\ \citenamefont
  {Riotto}}]{Apreda:2001us}%
  \BibitemOpen
  \bibfield  {author} {\bibinfo {author} {\bibfnamefont {R.}~\bibnamefont
  {Apreda}}, \bibinfo {author} {\bibfnamefont {M.}~\bibnamefont {Maggiore}},
  \bibinfo {author} {\bibfnamefont {A.}~\bibnamefont {Nicolis}}, \ and\
  \bibinfo {author} {\bibfnamefont {A.}~\bibnamefont {Riotto}},\ }\href
  {\doibase 10.1016/S0550-3213(02)00264-X} {\bibfield  {journal} {\bibinfo
  {journal} {Nucl. Phys. B}\ }\textbf {\bibinfo {volume} {631}},\ \bibinfo
  {pages} {342} (\bibinfo {year} {2002})},\ \Eprint
  {http://arxiv.org/abs/gr-qc/0107033} {arXiv:gr-qc/0107033} \BibitemShut
  {NoStop}%
\bibitem [{\citenamefont {Grojean}\ and\ \citenamefont
  {Servant}(2007)}]{Grojean:2006bp}%
  \BibitemOpen
  \bibfield  {author} {\bibinfo {author} {\bibfnamefont {C.}~\bibnamefont
  {Grojean}}\ and\ \bibinfo {author} {\bibfnamefont {G.}~\bibnamefont
  {Servant}},\ }\href {\doibase 10.1103/PhysRevD.75.043507} {\bibfield
  {journal} {\bibinfo  {journal} {Phys. Rev. D}\ }\textbf {\bibinfo {volume}
  {75}},\ \bibinfo {pages} {043507} (\bibinfo {year} {2007})},\ \Eprint
  {http://arxiv.org/abs/hep-ph/0607107} {arXiv:hep-ph/0607107} \BibitemShut
  {NoStop}%
\bibitem [{\citenamefont {Kakizaki}\ \emph {et~al.}(2015)\citenamefont
  {Kakizaki}, \citenamefont {Kanemura},\ and\ \citenamefont
  {Matsui}}]{Kakizaki:2015wua}%
  \BibitemOpen
  \bibfield  {author} {\bibinfo {author} {\bibfnamefont {M.}~\bibnamefont
  {Kakizaki}}, \bibinfo {author} {\bibfnamefont {S.}~\bibnamefont {Kanemura}},
  \ and\ \bibinfo {author} {\bibfnamefont {T.}~\bibnamefont {Matsui}},\ }\href
  {\doibase 10.1103/PhysRevD.92.115007} {\bibfield  {journal} {\bibinfo
  {journal} {Phys. Rev. D}\ }\textbf {\bibinfo {volume} {92}},\ \bibinfo
  {pages} {115007} (\bibinfo {year} {2015})},\ \Eprint
  {http://arxiv.org/abs/1509.08394} {arXiv:1509.08394 [hep-ph]} \BibitemShut
  {NoStop}%
\bibitem [{\citenamefont {Ellis}\ \emph {et~al.}(2019)\citenamefont {Ellis},
  \citenamefont {Lewicki},\ and\ \citenamefont {No}}]{Ellis:2018mja}%
  \BibitemOpen
  \bibfield  {author} {\bibinfo {author} {\bibfnamefont {J.}~\bibnamefont
  {Ellis}}, \bibinfo {author} {\bibfnamefont {M.}~\bibnamefont {Lewicki}}, \
  and\ \bibinfo {author} {\bibfnamefont {J.~M.}\ \bibnamefont {No}},\ }\href
  {\doibase 10.1088/1475-7516/2019/04/003} {\bibfield  {journal} {\bibinfo
  {journal} {JCAP}\ }\textbf {\bibinfo {volume} {04}},\ \bibinfo {pages} {003}
  (\bibinfo {year} {2019})},\ \Eprint {http://arxiv.org/abs/1809.08242}
  {arXiv:1809.08242 [hep-ph]} \BibitemShut {NoStop}%
\bibitem [{LIG()}]{LIGO}%
  \BibitemOpen
  \href@noop {} {\enquote {\bibinfo {title} {\uppercase{LIGO}},}\ }\bibinfo
  {howpublished} {https://www.ligo.caltech.edu/page/study-work}\BibitemShut
  {NoStop}%
\bibitem [{Vir()}]{Virgo}%
  \BibitemOpen
  \href@noop {} {\enquote {\bibinfo {title} {Virgo},}\ }\bibinfo {howpublished}
  {http://www.virgo-gw.eu/}\BibitemShut {NoStop}%
\bibitem [{LIS()}]{LISA}%
  \BibitemOpen
  \href@noop {} {\enquote {\bibinfo {title} {\uppercase{LISA}},}\ }\bibinfo
  {howpublished} {https://lisa.nasa.gov}\BibitemShut {NoStop}%
\bibitem [{DEC()}]{DECIGO}%
  \BibitemOpen
  \href@noop {} {\enquote {\bibinfo {title} {\uppercase{DECIGO}},}\ }\bibinfo
  {howpublished} {https://decigo.jp}\BibitemShut {NoStop}%
\bibitem [{\citenamefont {Craig}\ \emph {et~al.}(2020)\citenamefont {Craig},
  \citenamefont {Levi}, \citenamefont {Mariotti},\ and\ \citenamefont
  {Redigolo}}]{Craig:2020jfv}%
  \BibitemOpen
  \bibfield  {author} {\bibinfo {author} {\bibfnamefont {N.}~\bibnamefont
  {Craig}}, \bibinfo {author} {\bibfnamefont {N.}~\bibnamefont {Levi}},
  \bibinfo {author} {\bibfnamefont {A.}~\bibnamefont {Mariotti}}, \ and\
  \bibinfo {author} {\bibfnamefont {D.}~\bibnamefont {Redigolo}},\ }\href
  {\doibase 10.1007/JHEP02(2021)184} {\bibfield  {journal} {\bibinfo  {journal}
  {JHEP}\ }\textbf {\bibinfo {volume} {21}},\ \bibinfo {pages} {184} (\bibinfo
  {year} {2020})},\ \Eprint {http://arxiv.org/abs/2011.13949} {arXiv:2011.13949
  [hep-ph]} \BibitemShut {NoStop}%
\bibitem [{\citenamefont {Nelson}\ and\ \citenamefont
  {Seiberg}(1994)}]{Nelson:1993nf}%
  \BibitemOpen
  \bibfield  {author} {\bibinfo {author} {\bibfnamefont {A.~E.}\ \bibnamefont
  {Nelson}}\ and\ \bibinfo {author} {\bibfnamefont {N.}~\bibnamefont
  {Seiberg}},\ }\href {\doibase 10.1016/0550-3213(94)90577-0} {\bibfield
  {journal} {\bibinfo  {journal} {Nucl. Phys. B}\ }\textbf {\bibinfo {volume}
  {416}},\ \bibinfo {pages} {46} (\bibinfo {year} {1994})},\ \Eprint
  {http://arxiv.org/abs/hep-ph/9309299} {arXiv:hep-ph/9309299} \BibitemShut
  {NoStop}%
\bibitem [{\citenamefont {Intriligator}\ \emph {et~al.}(2006)\citenamefont
  {Intriligator}, \citenamefont {Seiberg},\ and\ \citenamefont
  {Shih}}]{Intriligator:2006dd}%
  \BibitemOpen
  \bibfield  {author} {\bibinfo {author} {\bibfnamefont {K.~A.}\ \bibnamefont
  {Intriligator}}, \bibinfo {author} {\bibfnamefont {N.}~\bibnamefont
  {Seiberg}}, \ and\ \bibinfo {author} {\bibfnamefont {D.}~\bibnamefont
  {Shih}},\ }\href {\doibase 10.1088/1126-6708/2006/04/021} {\bibfield
  {journal} {\bibinfo  {journal} {JHEP}\ }\textbf {\bibinfo {volume} {04}},\
  \bibinfo {pages} {021} (\bibinfo {year} {2006})},\ \Eprint
  {http://arxiv.org/abs/hep-th/0602239} {arXiv:hep-th/0602239} \BibitemShut
  {NoStop}%
\bibitem [{\citenamefont {Witten}(1982)}]{Witten:1982df}%
  \BibitemOpen
  \bibfield  {author} {\bibinfo {author} {\bibfnamefont {E.}~\bibnamefont
  {Witten}},\ }\href {\doibase 10.1016/0550-3213(82)90071-2} {\bibfield
  {journal} {\bibinfo  {journal} {Nucl. Phys. B}\ }\textbf {\bibinfo {volume}
  {202}},\ \bibinfo {pages} {253} (\bibinfo {year} {1982})}\BibitemShut
  {NoStop}%
\bibitem [{\citenamefont {Abel}\ \emph
  {et~al.}(2007{\natexlab{a}})\citenamefont {Abel}, \citenamefont {Chu},
  \citenamefont {Jaeckel},\ and\ \citenamefont {Khoze}}]{Abel:2006cr}%
  \BibitemOpen
  \bibfield  {author} {\bibinfo {author} {\bibfnamefont {S.~A.}\ \bibnamefont
  {Abel}}, \bibinfo {author} {\bibfnamefont {C.-S.}\ \bibnamefont {Chu}},
  \bibinfo {author} {\bibfnamefont {J.}~\bibnamefont {Jaeckel}}, \ and\
  \bibinfo {author} {\bibfnamefont {V.~V.}\ \bibnamefont {Khoze}},\ }\href
  {\doibase 10.1088/1126-6708/2007/01/089} {\bibfield  {journal} {\bibinfo
  {journal} {JHEP}\ }\textbf {\bibinfo {volume} {01}},\ \bibinfo {pages} {089}
  (\bibinfo {year} {2007}{\natexlab{a}})},\ \Eprint
  {http://arxiv.org/abs/hep-th/0610334} {arXiv:hep-th/0610334} \BibitemShut
  {NoStop}%
\bibitem [{\citenamefont {Abel}\ \emph
  {et~al.}(2007{\natexlab{b}})\citenamefont {Abel}, \citenamefont {Jaeckel},\
  and\ \citenamefont {Khoze}}]{Abel:2006my}%
  \BibitemOpen
  \bibfield  {author} {\bibinfo {author} {\bibfnamefont {S.~A.}\ \bibnamefont
  {Abel}}, \bibinfo {author} {\bibfnamefont {J.}~\bibnamefont {Jaeckel}}, \
  and\ \bibinfo {author} {\bibfnamefont {V.~V.}\ \bibnamefont {Khoze}},\ }\href
  {\doibase 10.1088/1126-6708/2007/01/015} {\bibfield  {journal} {\bibinfo
  {journal} {JHEP}\ }\textbf {\bibinfo {volume} {01}},\ \bibinfo {pages} {015}
  (\bibinfo {year} {2007}{\natexlab{b}})},\ \Eprint
  {http://arxiv.org/abs/hep-th/0611130} {arXiv:hep-th/0611130} \BibitemShut
  {NoStop}%
\bibitem [{\citenamefont {Fischler}\ \emph {et~al.}(2007)\citenamefont
  {Fischler}, \citenamefont {Kaplunovsky}, \citenamefont {Krishnan},
  \citenamefont {Mannelli},\ and\ \citenamefont {Torres}}]{Fischler:2006xh}%
  \BibitemOpen
  \bibfield  {author} {\bibinfo {author} {\bibfnamefont {W.}~\bibnamefont
  {Fischler}}, \bibinfo {author} {\bibfnamefont {V.}~\bibnamefont
  {Kaplunovsky}}, \bibinfo {author} {\bibfnamefont {C.}~\bibnamefont
  {Krishnan}}, \bibinfo {author} {\bibfnamefont {L.}~\bibnamefont {Mannelli}},
  \ and\ \bibinfo {author} {\bibfnamefont {M.~A.~C.}\ \bibnamefont {Torres}},\
  }\href {\doibase 10.1088/1126-6708/2007/03/107} {\bibfield  {journal}
  {\bibinfo  {journal} {JHEP}\ }\textbf {\bibinfo {volume} {03}},\ \bibinfo
  {pages} {107} (\bibinfo {year} {2007})},\ \Eprint
  {http://arxiv.org/abs/hep-th/0611018} {arXiv:hep-th/0611018} \BibitemShut
  {NoStop}%
\bibitem [{\citenamefont {Craig}\ \emph {et~al.}(2007)\citenamefont {Craig},
  \citenamefont {Fox},\ and\ \citenamefont {Wacker}}]{Craig:2006kx}%
  \BibitemOpen
  \bibfield  {author} {\bibinfo {author} {\bibfnamefont {N.~J.}\ \bibnamefont
  {Craig}}, \bibinfo {author} {\bibfnamefont {P.~J.}\ \bibnamefont {Fox}}, \
  and\ \bibinfo {author} {\bibfnamefont {J.~G.}\ \bibnamefont {Wacker}},\
  }\href {\doibase 10.1103/PhysRevD.75.085006} {\bibfield  {journal} {\bibinfo
  {journal} {Phys. Rev. D}\ }\textbf {\bibinfo {volume} {75}},\ \bibinfo
  {pages} {085006} (\bibinfo {year} {2007})},\ \Eprint
  {http://arxiv.org/abs/hep-th/0611006} {arXiv:hep-th/0611006} \BibitemShut
  {NoStop}%
\bibitem [{\citenamefont {Weinberg}(1982)}]{Weinberg:1982zq}%
  \BibitemOpen
  \bibfield  {author} {\bibinfo {author} {\bibfnamefont {S.}~\bibnamefont
  {Weinberg}},\ }\href {\doibase 10.1103/PhysRevLett.48.1303} {\bibfield
  {journal} {\bibinfo  {journal} {Phys. Rev. Lett.}\ }\textbf {\bibinfo
  {volume} {48}},\ \bibinfo {pages} {1303} (\bibinfo {year}
  {1982})}\BibitemShut {NoStop}%
\bibitem [{\citenamefont {Hook}\ \emph {et~al.}(2018)\citenamefont {Hook},
  \citenamefont {McGehee},\ and\ \citenamefont {Murayama}}]{Hook:2018sai}%
  \BibitemOpen
  \bibfield  {author} {\bibinfo {author} {\bibfnamefont {A.}~\bibnamefont
  {Hook}}, \bibinfo {author} {\bibfnamefont {R.}~\bibnamefont {McGehee}}, \
  and\ \bibinfo {author} {\bibfnamefont {H.}~\bibnamefont {Murayama}},\ }\href
  {\doibase 10.1103/PhysRevD.98.115036} {\bibfield  {journal} {\bibinfo
  {journal} {Phys. Rev. D}\ }\textbf {\bibinfo {volume} {98}},\ \bibinfo
  {pages} {115036} (\bibinfo {year} {2018})},\ \Eprint
  {http://arxiv.org/abs/1801.10160} {arXiv:1801.10160 [hep-ph]} \BibitemShut
  {NoStop}%
\bibitem [{\citenamefont {Seiberg}(1995)}]{Seiberg:1994pq}%
  \BibitemOpen
  \bibfield  {author} {\bibinfo {author} {\bibfnamefont {N.}~\bibnamefont
  {Seiberg}},\ }\href {\doibase 10.1016/0550-3213(94)00023-8} {\bibfield
  {journal} {\bibinfo  {journal} {Nucl. Phys. B}\ }\textbf {\bibinfo {volume}
  {435}},\ \bibinfo {pages} {129} (\bibinfo {year} {1995})},\ \Eprint
  {http://arxiv.org/abs/hep-th/9411149} {arXiv:hep-th/9411149} \BibitemShut
  {NoStop}%
\bibitem [{\citenamefont {Intriligator}\ and\ \citenamefont
  {Seiberg}(1996)}]{Intriligator:1995au}%
  \BibitemOpen
  \bibfield  {author} {\bibinfo {author} {\bibfnamefont {K.~A.}\ \bibnamefont
  {Intriligator}}\ and\ \bibinfo {author} {\bibfnamefont {N.}~\bibnamefont
  {Seiberg}},\ }\href {\doibase 10.1016/0920-5632(95)00626-5} {\bibfield
  {journal} {\bibinfo  {journal} {Nucl. Phys. B Proc. Suppl.}\ }\textbf
  {\bibinfo {volume} {45BC}},\ \bibinfo {pages} {1} (\bibinfo {year} {1996})},\
  \Eprint {http://arxiv.org/abs/hep-th/9509066} {arXiv:hep-th/9509066}
  \BibitemShut {NoStop}%
\bibitem [{\citenamefont {Peskin}(1997)}]{Peskin:1997qi}%
  \BibitemOpen
  \bibfield  {author} {\bibinfo {author} {\bibfnamefont {M.~E.}\ \bibnamefont
  {Peskin}}\ }(\bibinfo {year} {1997})\ pp.\ \bibinfo {pages} {729--809},\
  \Eprint {http://arxiv.org/abs/hep-th/9702094} {arXiv:hep-th/9702094}
  \BibitemShut {NoStop}%
\bibitem [{\citenamefont {Coleman}\ and\ \citenamefont
  {Weinberg}(1973)}]{Coleman:1973jx}%
  \BibitemOpen
  \bibfield  {author} {\bibinfo {author} {\bibfnamefont {S.~R.}\ \bibnamefont
  {Coleman}}\ and\ \bibinfo {author} {\bibfnamefont {E.~J.}\ \bibnamefont
  {Weinberg}},\ }\href {\doibase 10.1103/PhysRevD.7.1888} {\bibfield  {journal}
  {\bibinfo  {journal} {Phys. Rev. D}\ }\textbf {\bibinfo {volume} {7}},\
  \bibinfo {pages} {1888} (\bibinfo {year} {1973})}\BibitemShut {NoStop}%
\bibitem [{\citenamefont {Intriligator}\ and\ \citenamefont
  {Seiberg}(2007)}]{Intriligator:2007cp}%
  \BibitemOpen
  \bibfield  {author} {\bibinfo {author} {\bibfnamefont {K.~A.}\ \bibnamefont
  {Intriligator}}\ and\ \bibinfo {author} {\bibfnamefont {N.}~\bibnamefont
  {Seiberg}},\ }\href {\doibase 10.1088/0264-9381/24/21/S02} {\bibfield
  {journal} {\bibinfo  {journal} {Class. Quant. Grav.}\ }\textbf {\bibinfo
  {volume} {24}},\ \bibinfo {pages} {S741} (\bibinfo {year} {2007})},\ \Eprint
  {http://arxiv.org/abs/hep-ph/0702069} {arXiv:hep-ph/0702069} \BibitemShut
  {NoStop}%
\bibitem [{\citenamefont {Jackiw}(1974)}]{Jackiw:1974cv}%
  \BibitemOpen
  \bibfield  {author} {\bibinfo {author} {\bibfnamefont {R.}~\bibnamefont
  {Jackiw}},\ }\href {\doibase 10.1103/PhysRevD.9.1686} {\bibfield  {journal}
  {\bibinfo  {journal} {Phys. Rev. D}\ }\textbf {\bibinfo {volume} {9}},\
  \bibinfo {pages} {1686} (\bibinfo {year} {1974})}\BibitemShut {NoStop}%
\bibitem [{\citenamefont {Weinberg}(1974)}]{Weinberg:1974hy}%
  \BibitemOpen
  \bibfield  {author} {\bibinfo {author} {\bibfnamefont {S.}~\bibnamefont
  {Weinberg}},\ }\href {\doibase 10.1103/PhysRevD.9.3357} {\bibfield  {journal}
  {\bibinfo  {journal} {Phys. Rev. D}\ }\textbf {\bibinfo {volume} {9}},\
  \bibinfo {pages} {3357} (\bibinfo {year} {1974})}\BibitemShut {NoStop}%
\bibitem [{\citenamefont {Martin}(1998)}]{Martin:1997ns}%
  \BibitemOpen
  \bibfield  {author} {\bibinfo {author} {\bibfnamefont {S.~P.}\ \bibnamefont
  {Martin}},\ }\href {\doibase 10.1142/9789812839657_0001} {\bibfield
  {journal} {\bibinfo  {journal} {Adv. Ser. Direct. High Energy Phys.}\
  }\textbf {\bibinfo {volume} {18}},\ \bibinfo {pages} {1} (\bibinfo {year}
  {1998})},\ \Eprint {http://arxiv.org/abs/hep-ph/9709356}
  {arXiv:hep-ph/9709356} \BibitemShut {NoStop}%
\bibitem [{\citenamefont {Osato}\ \emph {et~al.}(2016)\citenamefont {Osato},
  \citenamefont {Sekiguchi}, \citenamefont {Shirasaki}, \citenamefont
  {Kamada},\ and\ \citenamefont {Yoshida}}]{Osato:2016ixc}%
  \BibitemOpen
  \bibfield  {author} {\bibinfo {author} {\bibfnamefont {K.}~\bibnamefont
  {Osato}}, \bibinfo {author} {\bibfnamefont {T.}~\bibnamefont {Sekiguchi}},
  \bibinfo {author} {\bibfnamefont {M.}~\bibnamefont {Shirasaki}}, \bibinfo
  {author} {\bibfnamefont {A.}~\bibnamefont {Kamada}}, \ and\ \bibinfo {author}
  {\bibfnamefont {N.}~\bibnamefont {Yoshida}},\ }\href {\doibase
  10.1088/1475-7516/2016/06/004} {\bibfield  {journal} {\bibinfo  {journal}
  {JCAP}\ }\textbf {\bibinfo {volume} {06}},\ \bibinfo {pages} {004} (\bibinfo
  {year} {2016})},\ \Eprint {http://arxiv.org/abs/1601.07386} {arXiv:1601.07386
  [astro-ph.CO]} \BibitemShut {NoStop}%
\bibitem [{\citenamefont {Viel}\ \emph {et~al.}(2005)\citenamefont {Viel},
  \citenamefont {Lesgourgues}, \citenamefont {Haehnelt}, \citenamefont
  {Matarrese},\ and\ \citenamefont {Riotto}}]{Viel:2005qj}%
  \BibitemOpen
  \bibfield  {author} {\bibinfo {author} {\bibfnamefont {M.}~\bibnamefont
  {Viel}}, \bibinfo {author} {\bibfnamefont {J.}~\bibnamefont {Lesgourgues}},
  \bibinfo {author} {\bibfnamefont {M.~G.}\ \bibnamefont {Haehnelt}}, \bibinfo
  {author} {\bibfnamefont {S.}~\bibnamefont {Matarrese}}, \ and\ \bibinfo
  {author} {\bibfnamefont {A.}~\bibnamefont {Riotto}},\ }\href {\doibase
  10.1103/PhysRevD.71.063534} {\bibfield  {journal} {\bibinfo  {journal} {Phys.
  Rev. D}\ }\textbf {\bibinfo {volume} {71}},\ \bibinfo {pages} {063534}
  (\bibinfo {year} {2005})},\ \Eprint {http://arxiv.org/abs/astro-ph/0501562}
  {arXiv:astro-ph/0501562} \BibitemShut {NoStop}%
\bibitem [{\citenamefont {Kawasaki}\ \emph {et~al.}(2008)\citenamefont
  {Kawasaki}, \citenamefont {Kohri}, \citenamefont {Moroi},\ and\ \citenamefont
  {Yotsuyanagi}}]{Kawasaki:2008qe}%
  \BibitemOpen
  \bibfield  {author} {\bibinfo {author} {\bibfnamefont {M.}~\bibnamefont
  {Kawasaki}}, \bibinfo {author} {\bibfnamefont {K.}~\bibnamefont {Kohri}},
  \bibinfo {author} {\bibfnamefont {T.}~\bibnamefont {Moroi}}, \ and\ \bibinfo
  {author} {\bibfnamefont {A.}~\bibnamefont {Yotsuyanagi}},\ }\href {\doibase
  10.1103/PhysRevD.78.065011} {\bibfield  {journal} {\bibinfo  {journal} {Phys.
  Rev. D}\ }\textbf {\bibinfo {volume} {78}},\ \bibinfo {pages} {065011}
  (\bibinfo {year} {2008})},\ \Eprint {http://arxiv.org/abs/0804.3745}
  {arXiv:0804.3745 [hep-ph]} \BibitemShut {NoStop}%
\bibitem [{\citenamefont {Moroi}\ \emph {et~al.}(1993)\citenamefont {Moroi},
  \citenamefont {Murayama},\ and\ \citenamefont {Yamaguchi}}]{Moroi:1993mb}%
  \BibitemOpen
  \bibfield  {author} {\bibinfo {author} {\bibfnamefont {T.}~\bibnamefont
  {Moroi}}, \bibinfo {author} {\bibfnamefont {H.}~\bibnamefont {Murayama}}, \
  and\ \bibinfo {author} {\bibfnamefont {M.}~\bibnamefont {Yamaguchi}},\ }\href
  {\doibase 10.1016/0370-2693(93)91434-O} {\bibfield  {journal} {\bibinfo
  {journal} {Phys. Lett. B}\ }\textbf {\bibinfo {volume} {303}},\ \bibinfo
  {pages} {289} (\bibinfo {year} {1993})}\BibitemShut {NoStop}%
\bibitem [{\citenamefont {Cyburt}\ \emph {et~al.}(2016)\citenamefont {Cyburt},
  \citenamefont {Fields}, \citenamefont {Olive},\ and\ \citenamefont
  {Yeh}}]{Cyburt:2015mya}%
  \BibitemOpen
  \bibfield  {author} {\bibinfo {author} {\bibfnamefont {R.~H.}\ \bibnamefont
  {Cyburt}}, \bibinfo {author} {\bibfnamefont {B.~D.}\ \bibnamefont {Fields}},
  \bibinfo {author} {\bibfnamefont {K.~A.}\ \bibnamefont {Olive}}, \ and\
  \bibinfo {author} {\bibfnamefont {T.-H.}\ \bibnamefont {Yeh}},\ }\href
  {\doibase 10.1103/RevModPhys.88.015004} {\bibfield  {journal} {\bibinfo
  {journal} {Rev. Mod. Phys.}\ }\textbf {\bibinfo {volume} {88}},\ \bibinfo
  {pages} {015004} (\bibinfo {year} {2016})},\ \Eprint
  {http://arxiv.org/abs/1505.01076} {arXiv:1505.01076 [astro-ph.CO]}
  \BibitemShut {NoStop}%
\bibitem [{\citenamefont {Murayama}\ and\ \citenamefont
  {Nomura}(2007)}]{Murayama:2006yf}%
  \BibitemOpen
  \bibfield  {author} {\bibinfo {author} {\bibfnamefont {H.}~\bibnamefont
  {Murayama}}\ and\ \bibinfo {author} {\bibfnamefont {Y.}~\bibnamefont
  {Nomura}},\ }\href {\doibase 10.1103/PhysRevLett.98.151803} {\bibfield
  {journal} {\bibinfo  {journal} {Phys. Rev. Lett.}\ }\textbf {\bibinfo
  {volume} {98}},\ \bibinfo {pages} {151803} (\bibinfo {year} {2007})},\
  \Eprint {http://arxiv.org/abs/hep-ph/0612186} {arXiv:hep-ph/0612186}
  \BibitemShut {NoStop}%
\bibitem [{\citenamefont {Baer}\ \emph {et~al.}(2020)\citenamefont {Baer},
  \citenamefont {Barger}, \citenamefont {Salam}, \citenamefont {Sengupta},\
  and\ \citenamefont {Sinha}}]{Baer:2020kwz}%
  \BibitemOpen
  \bibfield  {author} {\bibinfo {author} {\bibfnamefont {H.}~\bibnamefont
  {Baer}}, \bibinfo {author} {\bibfnamefont {V.}~\bibnamefont {Barger}},
  \bibinfo {author} {\bibfnamefont {S.}~\bibnamefont {Salam}}, \bibinfo
  {author} {\bibfnamefont {D.}~\bibnamefont {Sengupta}}, \ and\ \bibinfo
  {author} {\bibfnamefont {K.}~\bibnamefont {Sinha}},\ }\href {\doibase
  10.1140/epjst/e2020-000020-x} {\bibfield  {journal} {\bibinfo  {journal}
  {Eur. Phys. J. ST}\ }\textbf {\bibinfo {volume} {229}},\ \bibinfo {pages}
  {3085} (\bibinfo {year} {2020})},\ \Eprint {http://arxiv.org/abs/2002.03013}
  {arXiv:2002.03013 [hep-ph]} \BibitemShut {NoStop}%
\bibitem [{\citenamefont {Yoshimatsu}(2019)}]{Yoshimatsu:2019zfv}%
  \BibitemOpen
  \bibfield  {author} {\bibinfo {author} {\bibfnamefont {N.}~\bibnamefont
  {Yoshimatsu}},\ }\href {\doibase 10.1103/PhysRevD.100.096017} {\bibfield
  {journal} {\bibinfo  {journal} {Phys. Rev. D}\ }\textbf {\bibinfo {volume}
  {100}},\ \bibinfo {pages} {096017} (\bibinfo {year} {2019})},\ \Eprint
  {http://arxiv.org/abs/1911.04907} {arXiv:1911.04907 [hep-ph]} \BibitemShut
  {NoStop}%
\bibitem [{\citenamefont {Meade}\ \emph {et~al.}(2009)\citenamefont {Meade},
  \citenamefont {Seiberg},\ and\ \citenamefont {Shih}}]{Meade:2008wd}%
  \BibitemOpen
  \bibfield  {author} {\bibinfo {author} {\bibfnamefont {P.}~\bibnamefont
  {Meade}}, \bibinfo {author} {\bibfnamefont {N.}~\bibnamefont {Seiberg}}, \
  and\ \bibinfo {author} {\bibfnamefont {D.}~\bibnamefont {Shih}},\ }\href
  {\doibase 10.1143/PTPS.177.143} {\bibfield  {journal} {\bibinfo  {journal}
  {Prog. Theor. Phys. Suppl.}\ }\textbf {\bibinfo {volume} {177}},\ \bibinfo
  {pages} {143} (\bibinfo {year} {2009})},\ \Eprint
  {http://arxiv.org/abs/0801.3278} {arXiv:0801.3278 [hep-ph]} \BibitemShut
  {NoStop}%
\bibitem [{\citenamefont {Carpenter}\ \emph {et~al.}(2009)\citenamefont
  {Carpenter}, \citenamefont {Dine}, \citenamefont {Festuccia},\ and\
  \citenamefont {Mason}}]{Carpenter:2008wi}%
  \BibitemOpen
  \bibfield  {author} {\bibinfo {author} {\bibfnamefont {L.~M.}\ \bibnamefont
  {Carpenter}}, \bibinfo {author} {\bibfnamefont {M.}~\bibnamefont {Dine}},
  \bibinfo {author} {\bibfnamefont {G.}~\bibnamefont {Festuccia}}, \ and\
  \bibinfo {author} {\bibfnamefont {J.~D.}\ \bibnamefont {Mason}},\ }\href
  {\doibase 10.1103/PhysRevD.79.035002} {\bibfield  {journal} {\bibinfo
  {journal} {Phys. Rev. D}\ }\textbf {\bibinfo {volume} {79}},\ \bibinfo
  {pages} {035002} (\bibinfo {year} {2009})},\ \Eprint
  {http://arxiv.org/abs/0805.2944} {arXiv:0805.2944 [hep-ph]} \BibitemShut
  {NoStop}%
\bibitem [{\citenamefont {Turner}\ and\ \citenamefont
  {Wilczek}(1990)}]{Turner:1990rc}%
  \BibitemOpen
  \bibfield  {author} {\bibinfo {author} {\bibfnamefont {M.~S.}\ \bibnamefont
  {Turner}}\ and\ \bibinfo {author} {\bibfnamefont {F.}~\bibnamefont
  {Wilczek}},\ }\href {\doibase 10.1103/PhysRevLett.65.3080} {\bibfield
  {journal} {\bibinfo  {journal} {Phys. Rev. Lett.}\ }\textbf {\bibinfo
  {volume} {65}},\ \bibinfo {pages} {3080} (\bibinfo {year}
  {1990})}\BibitemShut {NoStop}%
\bibitem [{\citenamefont {Hogan}(1986)}]{Hogan:1986qda}%
  \BibitemOpen
  \bibfield  {author} {\bibinfo {author} {\bibfnamefont {C.~J.}\ \bibnamefont
  {Hogan}},\ }\href@noop {} {\bibfield  {journal} {\bibinfo  {journal} {Mon.
  Not. Roy. Astron. Soc.}\ }\textbf {\bibinfo {volume} {218}},\ \bibinfo
  {pages} {629} (\bibinfo {year} {1986})}\BibitemShut {NoStop}%
\bibitem [{\citenamefont {Kosowsky}\ \emph
  {et~al.}(1992{\natexlab{a}})\citenamefont {Kosowsky}, \citenamefont
  {Turner},\ and\ \citenamefont {Watkins}}]{Kosowsky:1992rz}%
  \BibitemOpen
  \bibfield  {author} {\bibinfo {author} {\bibfnamefont {A.}~\bibnamefont
  {Kosowsky}}, \bibinfo {author} {\bibfnamefont {M.~S.}\ \bibnamefont
  {Turner}}, \ and\ \bibinfo {author} {\bibfnamefont {R.}~\bibnamefont
  {Watkins}},\ }\href {\doibase 10.1103/PhysRevLett.69.2026} {\bibfield
  {journal} {\bibinfo  {journal} {Phys. Rev. Lett.}\ }\textbf {\bibinfo
  {volume} {69}},\ \bibinfo {pages} {2026} (\bibinfo {year}
  {1992}{\natexlab{a}})}\BibitemShut {NoStop}%
\bibitem [{\citenamefont {Kamionkowski}\ \emph {et~al.}(1994)\citenamefont
  {Kamionkowski}, \citenamefont {Kosowsky},\ and\ \citenamefont
  {Turner}}]{Kamionkowski:1993fg}%
  \BibitemOpen
  \bibfield  {author} {\bibinfo {author} {\bibfnamefont {M.}~\bibnamefont
  {Kamionkowski}}, \bibinfo {author} {\bibfnamefont {A.}~\bibnamefont
  {Kosowsky}}, \ and\ \bibinfo {author} {\bibfnamefont {M.~S.}\ \bibnamefont
  {Turner}},\ }\href {\doibase 10.1103/PhysRevD.49.2837} {\bibfield  {journal}
  {\bibinfo  {journal} {Phys. Rev. D}\ }\textbf {\bibinfo {volume} {49}},\
  \bibinfo {pages} {2837} (\bibinfo {year} {1994})},\ \Eprint
  {http://arxiv.org/abs/astro-ph/9310044} {arXiv:astro-ph/9310044} \BibitemShut
  {NoStop}%
\bibitem [{\citenamefont {Turner}\ \emph {et~al.}(1992)\citenamefont {Turner},
  \citenamefont {Weinberg},\ and\ \citenamefont {Widrow}}]{Turner:1992tz}%
  \BibitemOpen
  \bibfield  {author} {\bibinfo {author} {\bibfnamefont {M.~S.}\ \bibnamefont
  {Turner}}, \bibinfo {author} {\bibfnamefont {E.~J.}\ \bibnamefont
  {Weinberg}}, \ and\ \bibinfo {author} {\bibfnamefont {L.~M.}\ \bibnamefont
  {Widrow}},\ }\href {\doibase 10.1103/PhysRevD.46.2384} {\bibfield  {journal}
  {\bibinfo  {journal} {Phys. Rev. D}\ }\textbf {\bibinfo {volume} {46}},\
  \bibinfo {pages} {2384} (\bibinfo {year} {1992})}\BibitemShut {NoStop}%
\bibitem [{\citenamefont {Linde}(1983)}]{Linde:1981zj}%
  \BibitemOpen
  \bibfield  {author} {\bibinfo {author} {\bibfnamefont {A.~D.}\ \bibnamefont
  {Linde}},\ }\href {\doibase 10.1016/0550-3213(83)90072-X} {\bibfield
  {journal} {\bibinfo  {journal} {Nucl. Phys. B}\ }\textbf {\bibinfo {volume}
  {216}},\ \bibinfo {pages} {421} (\bibinfo {year} {1983})},\ \bibinfo {note}
  {[Erratum: Nucl.Phys.B 223, 544 (1983)]}\BibitemShut {NoStop}%
\bibitem [{\citenamefont {Coleman}\ \emph {et~al.}(1978)\citenamefont
  {Coleman}, \citenamefont {Glaser},\ and\ \citenamefont
  {Martin}}]{Coleman:1977th}%
  \BibitemOpen
  \bibfield  {author} {\bibinfo {author} {\bibfnamefont {S.~R.}\ \bibnamefont
  {Coleman}}, \bibinfo {author} {\bibfnamefont {V.}~\bibnamefont {Glaser}}, \
  and\ \bibinfo {author} {\bibfnamefont {A.}~\bibnamefont {Martin}},\ }\href
  {\doibase 10.1007/BF01609421} {\bibfield  {journal} {\bibinfo  {journal}
  {Commun. Math. Phys.}\ }\textbf {\bibinfo {volume} {58}},\ \bibinfo {pages}
  {211} (\bibinfo {year} {1978})}\BibitemShut {NoStop}%
\bibitem [{\citenamefont {Lee}\ and\ \citenamefont
  {Weinberg}(1986)}]{Lee:1985uv}%
  \BibitemOpen
  \bibfield  {author} {\bibinfo {author} {\bibfnamefont {K.-M.}\ \bibnamefont
  {Lee}}\ and\ \bibinfo {author} {\bibfnamefont {E.~J.}\ \bibnamefont
  {Weinberg}},\ }\href {\doibase 10.1016/0550-3213(86)90150-1} {\bibfield
  {journal} {\bibinfo  {journal} {Nucl. Phys. B}\ }\textbf {\bibinfo {volume}
  {267}},\ \bibinfo {pages} {181} (\bibinfo {year} {1986})}\BibitemShut
  {NoStop}%
\bibitem [{\citenamefont {Caprini}\ \emph {et~al.}(2016)\citenamefont {Caprini}
  \emph {et~al.}}]{Caprini:2015zlo}%
  \BibitemOpen
  \bibfield  {author} {\bibinfo {author} {\bibfnamefont {C.}~\bibnamefont
  {Caprini}} \emph {et~al.},\ }\href {\doibase 10.1088/1475-7516/2016/04/001}
  {\bibfield  {journal} {\bibinfo  {journal} {JCAP}\ }\textbf {\bibinfo
  {volume} {04}},\ \bibinfo {pages} {001} (\bibinfo {year} {2016})},\ \Eprint
  {http://arxiv.org/abs/1512.06239} {arXiv:1512.06239 [astro-ph.CO]}
  \BibitemShut {NoStop}%
\bibitem [{\citenamefont {Espinosa}\ \emph {et~al.}(2010)\citenamefont
  {Espinosa}, \citenamefont {Konstandin}, \citenamefont {No},\ and\
  \citenamefont {Servant}}]{Espinosa:2010hh}%
  \BibitemOpen
  \bibfield  {author} {\bibinfo {author} {\bibfnamefont {J.~R.}\ \bibnamefont
  {Espinosa}}, \bibinfo {author} {\bibfnamefont {T.}~\bibnamefont
  {Konstandin}}, \bibinfo {author} {\bibfnamefont {J.~M.}\ \bibnamefont {No}},
  \ and\ \bibinfo {author} {\bibfnamefont {G.}~\bibnamefont {Servant}},\ }\href
  {\doibase 10.1088/1475-7516/2010/06/028} {\bibfield  {journal} {\bibinfo
  {journal} {JCAP}\ }\textbf {\bibinfo {volume} {06}},\ \bibinfo {pages} {028}
  (\bibinfo {year} {2010})},\ \Eprint {http://arxiv.org/abs/1004.4187}
  {arXiv:1004.4187 [hep-ph]} \BibitemShut {NoStop}%
\bibitem [{\citenamefont {Kosowsky}\ \emph
  {et~al.}(1992{\natexlab{b}})\citenamefont {Kosowsky}, \citenamefont
  {Turner},\ and\ \citenamefont {Watkins}}]{Kosowsky:1991ua}%
  \BibitemOpen
  \bibfield  {author} {\bibinfo {author} {\bibfnamefont {A.}~\bibnamefont
  {Kosowsky}}, \bibinfo {author} {\bibfnamefont {M.~S.}\ \bibnamefont
  {Turner}}, \ and\ \bibinfo {author} {\bibfnamefont {R.}~\bibnamefont
  {Watkins}},\ }\href {\doibase 10.1103/PhysRevD.45.4514} {\bibfield  {journal}
  {\bibinfo  {journal} {Phys. Rev. D}\ }\textbf {\bibinfo {volume} {45}},\
  \bibinfo {pages} {4514} (\bibinfo {year} {1992}{\natexlab{b}})}\BibitemShut
  {NoStop}%
\bibitem [{\citenamefont {Kosowsky}\ and\ \citenamefont
  {Turner}(1993)}]{Kosowsky:1992vn}%
  \BibitemOpen
  \bibfield  {author} {\bibinfo {author} {\bibfnamefont {A.}~\bibnamefont
  {Kosowsky}}\ and\ \bibinfo {author} {\bibfnamefont {M.~S.}\ \bibnamefont
  {Turner}},\ }\href {\doibase 10.1103/PhysRevD.47.4372} {\bibfield  {journal}
  {\bibinfo  {journal} {Phys. Rev. D}\ }\textbf {\bibinfo {volume} {47}},\
  \bibinfo {pages} {4372} (\bibinfo {year} {1993})},\ \Eprint
  {http://arxiv.org/abs/astro-ph/9211004} {arXiv:astro-ph/9211004} \BibitemShut
  {NoStop}%
\bibitem [{\citenamefont {Caprini}\ \emph {et~al.}(2008)\citenamefont
  {Caprini}, \citenamefont {Durrer},\ and\ \citenamefont
  {Servant}}]{Caprini:2007xq}%
  \BibitemOpen
  \bibfield  {author} {\bibinfo {author} {\bibfnamefont {C.}~\bibnamefont
  {Caprini}}, \bibinfo {author} {\bibfnamefont {R.}~\bibnamefont {Durrer}}, \
  and\ \bibinfo {author} {\bibfnamefont {G.}~\bibnamefont {Servant}},\ }\href
  {\doibase 10.1103/PhysRevD.77.124015} {\bibfield  {journal} {\bibinfo
  {journal} {Phys. Rev. D}\ }\textbf {\bibinfo {volume} {77}},\ \bibinfo
  {pages} {124015} (\bibinfo {year} {2008})},\ \Eprint
  {http://arxiv.org/abs/0711.2593} {arXiv:0711.2593 [astro-ph]} \BibitemShut
  {NoStop}%
\bibitem [{\citenamefont {Huber}\ and\ \citenamefont
  {Konstandin}(2008)}]{Huber:2008hg}%
  \BibitemOpen
  \bibfield  {author} {\bibinfo {author} {\bibfnamefont {S.~J.}\ \bibnamefont
  {Huber}}\ and\ \bibinfo {author} {\bibfnamefont {T.}~\bibnamefont
  {Konstandin}},\ }\href {\doibase 10.1088/1475-7516/2008/09/022} {\bibfield
  {journal} {\bibinfo  {journal} {JCAP}\ }\textbf {\bibinfo {volume} {09}},\
  \bibinfo {pages} {022} (\bibinfo {year} {2008})},\ \Eprint
  {http://arxiv.org/abs/0806.1828} {arXiv:0806.1828 [hep-ph]} \BibitemShut
  {NoStop}%
\bibitem [{\citenamefont {Hindmarsh}\ \emph {et~al.}(2014)\citenamefont
  {Hindmarsh}, \citenamefont {Huber}, \citenamefont {Rummukainen},\ and\
  \citenamefont {Weir}}]{Hindmarsh:2013xza}%
  \BibitemOpen
  \bibfield  {author} {\bibinfo {author} {\bibfnamefont {M.}~\bibnamefont
  {Hindmarsh}}, \bibinfo {author} {\bibfnamefont {S.~J.}\ \bibnamefont
  {Huber}}, \bibinfo {author} {\bibfnamefont {K.}~\bibnamefont {Rummukainen}},
  \ and\ \bibinfo {author} {\bibfnamefont {D.~J.}\ \bibnamefont {Weir}},\
  }\href {\doibase 10.1103/PhysRevLett.112.041301} {\bibfield  {journal}
  {\bibinfo  {journal} {Phys. Rev. Lett.}\ }\textbf {\bibinfo {volume} {112}},\
  \bibinfo {pages} {041301} (\bibinfo {year} {2014})},\ \Eprint
  {http://arxiv.org/abs/1304.2433} {arXiv:1304.2433 [hep-ph]} \BibitemShut
  {NoStop}%
\bibitem [{\citenamefont {Giblin}\ and\ \citenamefont
  {Mertens}(2013)}]{Giblin:2013kea}%
  \BibitemOpen
  \bibfield  {author} {\bibinfo {author} {\bibfnamefont {J.~T.}\ \bibnamefont
  {Giblin}, \bibfnamefont {Jr.}}\ and\ \bibinfo {author} {\bibfnamefont
  {J.~B.}\ \bibnamefont {Mertens}},\ }\href {\doibase 10.1007/JHEP12(2013)042}
  {\bibfield  {journal} {\bibinfo  {journal} {JHEP}\ }\textbf {\bibinfo
  {volume} {12}},\ \bibinfo {pages} {042} (\bibinfo {year} {2013})},\ \Eprint
  {http://arxiv.org/abs/1310.2948} {arXiv:1310.2948 [hep-th]} \BibitemShut
  {NoStop}%
\bibitem [{\citenamefont {Giblin}\ and\ \citenamefont
  {Mertens}(2014)}]{Giblin:2014qia}%
  \BibitemOpen
  \bibfield  {author} {\bibinfo {author} {\bibfnamefont {J.~T.}\ \bibnamefont
  {Giblin}}\ and\ \bibinfo {author} {\bibfnamefont {J.~B.}\ \bibnamefont
  {Mertens}},\ }\href {\doibase 10.1103/PhysRevD.90.023532} {\bibfield
  {journal} {\bibinfo  {journal} {Phys. Rev. D}\ }\textbf {\bibinfo {volume}
  {90}},\ \bibinfo {pages} {023532} (\bibinfo {year} {2014})},\ \Eprint
  {http://arxiv.org/abs/1405.4005} {arXiv:1405.4005 [astro-ph.CO]} \BibitemShut
  {NoStop}%
\bibitem [{\citenamefont {Hindmarsh}\ \emph {et~al.}(2015)\citenamefont
  {Hindmarsh}, \citenamefont {Huber}, \citenamefont {Rummukainen},\ and\
  \citenamefont {Weir}}]{Hindmarsh:2015qta}%
  \BibitemOpen
  \bibfield  {author} {\bibinfo {author} {\bibfnamefont {M.}~\bibnamefont
  {Hindmarsh}}, \bibinfo {author} {\bibfnamefont {S.~J.}\ \bibnamefont
  {Huber}}, \bibinfo {author} {\bibfnamefont {K.}~\bibnamefont {Rummukainen}},
  \ and\ \bibinfo {author} {\bibfnamefont {D.~J.}\ \bibnamefont {Weir}},\
  }\href {\doibase 10.1103/PhysRevD.92.123009} {\bibfield  {journal} {\bibinfo
  {journal} {Phys. Rev. D}\ }\textbf {\bibinfo {volume} {92}},\ \bibinfo
  {pages} {123009} (\bibinfo {year} {2015})},\ \Eprint
  {http://arxiv.org/abs/1504.03291} {arXiv:1504.03291 [astro-ph.CO]}
  \BibitemShut {NoStop}%
\bibitem [{\citenamefont {Caprini}\ and\ \citenamefont
  {Durrer}(2006)}]{Caprini:2006jb}%
  \BibitemOpen
  \bibfield  {author} {\bibinfo {author} {\bibfnamefont {C.}~\bibnamefont
  {Caprini}}\ and\ \bibinfo {author} {\bibfnamefont {R.}~\bibnamefont
  {Durrer}},\ }\href {\doibase 10.1103/PhysRevD.74.063521} {\bibfield
  {journal} {\bibinfo  {journal} {Phys. Rev. D}\ }\textbf {\bibinfo {volume}
  {74}},\ \bibinfo {pages} {063521} (\bibinfo {year} {2006})},\ \Eprint
  {http://arxiv.org/abs/astro-ph/0603476} {arXiv:astro-ph/0603476} \BibitemShut
  {NoStop}%
\bibitem [{\citenamefont {Kahniashvili}\ \emph
  {et~al.}(2008{\natexlab{a}})\citenamefont {Kahniashvili}, \citenamefont
  {Kosowsky}, \citenamefont {Gogoberidze},\ and\ \citenamefont
  {Maravin}}]{Kahniashvili:2008pf}%
  \BibitemOpen
  \bibfield  {author} {\bibinfo {author} {\bibfnamefont {T.}~\bibnamefont
  {Kahniashvili}}, \bibinfo {author} {\bibfnamefont {A.}~\bibnamefont
  {Kosowsky}}, \bibinfo {author} {\bibfnamefont {G.}~\bibnamefont
  {Gogoberidze}}, \ and\ \bibinfo {author} {\bibfnamefont {Y.}~\bibnamefont
  {Maravin}},\ }\href {\doibase 10.1103/PhysRevD.78.043003} {\bibfield
  {journal} {\bibinfo  {journal} {Phys. Rev. D}\ }\textbf {\bibinfo {volume}
  {78}},\ \bibinfo {pages} {043003} (\bibinfo {year} {2008}{\natexlab{a}})},\
  \Eprint {http://arxiv.org/abs/0806.0293} {arXiv:0806.0293 [astro-ph]}
  \BibitemShut {NoStop}%
\bibitem [{\citenamefont {Kahniashvili}\ \emph
  {et~al.}(2008{\natexlab{b}})\citenamefont {Kahniashvili}, \citenamefont
  {Campanelli}, \citenamefont {Gogoberidze}, \citenamefont {Maravin},\ and\
  \citenamefont {Ratra}}]{Kahniashvili:2008pe}%
  \BibitemOpen
  \bibfield  {author} {\bibinfo {author} {\bibfnamefont {T.}~\bibnamefont
  {Kahniashvili}}, \bibinfo {author} {\bibfnamefont {L.}~\bibnamefont
  {Campanelli}}, \bibinfo {author} {\bibfnamefont {G.}~\bibnamefont
  {Gogoberidze}}, \bibinfo {author} {\bibfnamefont {Y.}~\bibnamefont
  {Maravin}}, \ and\ \bibinfo {author} {\bibfnamefont {B.}~\bibnamefont
  {Ratra}},\ }\href {\doibase 10.1103/PhysRevD.78.123006} {\bibfield  {journal}
  {\bibinfo  {journal} {Phys. Rev. D}\ }\textbf {\bibinfo {volume} {78}},\
  \bibinfo {pages} {123006} (\bibinfo {year} {2008}{\natexlab{b}})},\ \bibinfo
  {note} {[Erratum: Phys.Rev.D 79, 109901 (2009)]},\ \Eprint
  {http://arxiv.org/abs/0809.1899} {arXiv:0809.1899 [astro-ph]} \BibitemShut
  {NoStop}%
\bibitem [{\citenamefont {Kahniashvili}\ \emph {et~al.}(2010)\citenamefont
  {Kahniashvili}, \citenamefont {Kisslinger},\ and\ \citenamefont
  {Stevens}}]{Kahniashvili:2009mf}%
  \BibitemOpen
  \bibfield  {author} {\bibinfo {author} {\bibfnamefont {T.}~\bibnamefont
  {Kahniashvili}}, \bibinfo {author} {\bibfnamefont {L.}~\bibnamefont
  {Kisslinger}}, \ and\ \bibinfo {author} {\bibfnamefont {T.}~\bibnamefont
  {Stevens}},\ }\href {\doibase 10.1103/PhysRevD.81.023004} {\bibfield
  {journal} {\bibinfo  {journal} {Phys. Rev. D}\ }\textbf {\bibinfo {volume}
  {81}},\ \bibinfo {pages} {023004} (\bibinfo {year} {2010})},\ \Eprint
  {http://arxiv.org/abs/0905.0643} {arXiv:0905.0643 [astro-ph.CO]} \BibitemShut
  {NoStop}%
\bibitem [{\citenamefont {Caprini}\ \emph {et~al.}(2009)\citenamefont
  {Caprini}, \citenamefont {Durrer},\ and\ \citenamefont
  {Servant}}]{Caprini:2009yp}%
  \BibitemOpen
  \bibfield  {author} {\bibinfo {author} {\bibfnamefont {C.}~\bibnamefont
  {Caprini}}, \bibinfo {author} {\bibfnamefont {R.}~\bibnamefont {Durrer}}, \
  and\ \bibinfo {author} {\bibfnamefont {G.}~\bibnamefont {Servant}},\ }\href
  {\doibase 10.1088/1475-7516/2009/12/024} {\bibfield  {journal} {\bibinfo
  {journal} {JCAP}\ }\textbf {\bibinfo {volume} {12}},\ \bibinfo {pages} {024}
  (\bibinfo {year} {2009})},\ \Eprint {http://arxiv.org/abs/0909.0622}
  {arXiv:0909.0622 [astro-ph.CO]} \BibitemShut {NoStop}%
\bibitem [{\citenamefont {Kawamura}\ \emph {et~al.}(2021)\citenamefont
  {Kawamura} \emph {et~al.}}]{Kawamura:2020pcg}%
  \BibitemOpen
  \bibfield  {author} {\bibinfo {author} {\bibfnamefont {S.}~\bibnamefont
  {Kawamura}} \emph {et~al.},\ }\href {\doibase 10.1093/ptep/ptab019}
  {\bibfield  {journal} {\bibinfo  {journal} {PTEP}\ }\textbf {\bibinfo
  {volume} {2021}},\ \bibinfo {pages} {05A105} (\bibinfo {year} {2021})},\
  \Eprint {http://arxiv.org/abs/2006.13545} {arXiv:2006.13545 [gr-qc]}
  \BibitemShut {NoStop}%
\bibitem [{\citenamefont {Craig}(2008)}]{Craig:2008tv}%
  \BibitemOpen
  \bibfield  {author} {\bibinfo {author} {\bibfnamefont {N.~J.}\ \bibnamefont
  {Craig}},\ }\href {\doibase 10.1088/1126-6708/2008/02/059} {\bibfield
  {journal} {\bibinfo  {journal} {JHEP}\ }\textbf {\bibinfo {volume} {02}},\
  \bibinfo {pages} {059} (\bibinfo {year} {2008})},\ \Eprint
  {http://arxiv.org/abs/0801.2157} {arXiv:0801.2157 [hep-th]} \BibitemShut
  {NoStop}%
\bibitem [{\citenamefont {Dalianis}\ and\ \citenamefont
  {Lalak}(2010)}]{Dalianis:2010yk}%
  \BibitemOpen
  \bibfield  {author} {\bibinfo {author} {\bibfnamefont {I.}~\bibnamefont
  {Dalianis}}\ and\ \bibinfo {author} {\bibfnamefont {Z.}~\bibnamefont
  {Lalak}},\ }\href {\doibase 10.1007/JHEP12(2010)045} {\bibfield  {journal}
  {\bibinfo  {journal} {JHEP}\ }\textbf {\bibinfo {volume} {12}},\ \bibinfo
  {pages} {045} (\bibinfo {year} {2010})},\ \Eprint
  {http://arxiv.org/abs/1001.4106} {arXiv:1001.4106 [hep-ph]} \BibitemShut
  {NoStop}%
\bibitem [{\citenamefont {Dalianis}\ and\ \citenamefont
  {Lalak}(2011)}]{Dalianis:2010pq}%
  \BibitemOpen
  \bibfield  {author} {\bibinfo {author} {\bibfnamefont {I.}~\bibnamefont
  {Dalianis}}\ and\ \bibinfo {author} {\bibfnamefont {Z.}~\bibnamefont
  {Lalak}},\ }\href {\doibase 10.1016/j.physletb.2011.02.010} {\bibfield
  {journal} {\bibinfo  {journal} {Phys. Lett. B}\ }\textbf {\bibinfo {volume}
  {697}},\ \bibinfo {pages} {385} (\bibinfo {year} {2011})},\ \Eprint
  {http://arxiv.org/abs/1012.3157} {arXiv:1012.3157 [hep-ph]} \BibitemShut
  {NoStop}%
\end{thebibliography}%

\end{document}